\documentclass[american,aps,pra,superscriptaddress,reprint,showpacs]{revtex4-1}
\usepackage[T1]{fontenc}
\usepackage[latin9]{inputenc}
\usepackage{babel}
\usepackage{amsthm}
\usepackage{amsmath}
\usepackage{bm}
\usepackage{tabularx}
\usepackage{amssymb}
\usepackage{graphics,graphicx}
\usepackage{tikz}
\usepackage{float}
\usepackage{subfigure}
\usepackage{mathrsfs}
\usepackage{epstopdf}
\usepackage{fancyhdr}

\usepackage[unicode=true,pdfusetitle, bookmarks=true,bookmarksnumbered=false,bookmarksopen=false,
 breaklinks=false,pdfborder={0 0 0},backref=false,colorlinks=false]
 {hyperref}
\hypersetup{colorlinks,linkcolor=myurlcolor,citecolor=myurlcolor,urlcolor=myurlcolor}

\makeatletter
\@ifundefined{textcolor}{}
{%
 \definecolor{BLACK}{gray}{0}
 \definecolor{WHITE}{gray}{1}
 \definecolor{RED}{rgb}{1,0,0}
 \definecolor{GREEN}{rgb}{0,1,0}
 \definecolor{BLUE}{rgb}{0,0,1}
 \definecolor{CYAN}{cmyk}{1,0,0,0}
 \definecolor{MAGENTA}{cmyk}{0,1,0,0}
 \definecolor{YELLOW}{cmyk}{0,0,1,0}
 }
\theoremstyle{plain}

\ifx\proof\undefined
\newenvironment{proof}[1][\protect\proofname]{\par
\normalfont\topsep6\p@\@plus6\p@\relax
\trivlist
\itemindent\parindent
\item[\hskip\labelsep
\scshape
#1]\ignorespaces
}{%
\endtrivlist\@endpefalse
}
\providecommand{\proofname}{Proof}
\fi

\usepackage{times}
\usepackage{txfonts}
\usepackage{braket}
\usepackage{colortbl}
\definecolor{myurlcolor}{rgb}{0,0,0.7}

\makeatother

\providecommand{\theoremname}{Theorem}

\usepackage{times}

\newcommand{\bq}{\begin{eqnarray}}
\newcommand{\eq}{\end{eqnarray}}
\newcommand{\be}{\begin{equation}}
\newcommand{\ee}{\end{equation}}
\newcommand{\ei}{\mathrm{e}}
\newcommand{\tr}{\operatorname{tr}}

\newcommand{\abs}[1]{\left| #1 \right|}

\begin{document}
\title{Gaussian interferometric power as a measure of
continuous variable non-Markovianity}

\begin{abstract}
We investigate the non-Markovianity of continuous variable Gaussian quantum channels through the evolution of an operational metrological quantifier, namely the Gaussian interferometric power,
which captures the minimal precision that can be achieved using bipartite Gaussian probes in a black-box phase estimation setup, where the phase shift generator is \emph{a priori} unknown.
We observe that the monotonicity of the Gaussian interferometric power under the action of local Gaussian quantum channels on the ancillary arm of the bipartite probes is a natural indicator of Markovian dynamics; consequently, its breakdown for specific maps can be used to construct a witness and an effective quantifier of non-Markovianity. In our work, we consider two paradigmatic Gaussian models, the damping master equation and the quantum Brownian motion, and identify analytically and numerically the parameter regimes that give rise to non-Markovian dynamics. We then quantify the degree of non-Markovianity of the channels in terms of Gaussian interferometric power, showing in particular that even nonentangled probes can be useful to witness non-Markovianity. This establishes an interesting link between the dynamics of bipartite continuous variable open systems and their potential for optical interferometry. The results are an important supplement to the recent research on characterization of non-Markovianity in continuous variable systems.
\end{abstract}


\author{Leonardo A. M. Souza}
\email{leonardoamsouza@ufv.br}
\affiliation{Universidade Federal de Vi{\c c}osa - Campus Florestal,
LMG818 Km6, Minas Gerais, Florestal 35690-000, Brazil}
\affiliation{$\mbox{School of Mathematical Sciences, The University of Nottingham, University Park,
Nottingham NG7 2RD, United Kingdom}$}

\author{Himadri Shekhar Dhar}
\affiliation{Harish-Chandra Research Institute, Chhatnag Road, Jhunsi, Allahabad 211 019, India}

\author{Manabendra Nath Bera}

\affiliation{ICFO-The Institute of Photonic Sciences, Mediterranean Technology
Park, 08860 Castelldefels (Barcelona), Spain}

\affiliation{Harish-Chandra Research Institute, Chhatnag Road, Jhunsi, Allahabad 211 019, India}

\author{Pietro Liuzzo-Scorpo}
\affiliation{$\mbox{School of Mathematical Sciences, The University of Nottingham, University Park,
Nottingham NG7 2RD, United Kingdom}$}

\author{Gerardo Adesso}
\affiliation{$\mbox{School of Mathematical Sciences, The University of Nottingham, University Park,
Nottingham NG7 2RD, United Kingdom}$}

\date{\today}

\pacs{03.65.Yz, 03.65.Ta, 42.50.Lc}

\maketitle

\section{Introduction}
The study of the dynamics of open quantum systems has received a lot of attention in recent research \cite{Petruccione, Weiss, Rivas-book}, especially with regards to the nature of the interaction between the system and its surrounding environment. This is a crucial aspect in quantum information theory, where quantum resources irreversibly \emph{decohere} under nonunitary evolutions described through quantum channels, and are rendered less useful for quantum protocols. In ideal terms, the dynamics of open systems can be defined via a weak system-environment coupling and a long system relaxation time-scale, that entails a non-retrievable transfer of information from the system to the environment. Such dynamics are modelled using master equations of the Lindblad form \cite{lind1976, gorini1976, carmichael}, and are called Markovian \cite{gorini1978}. However, a more realistic description of open systems dynamics incorporates a stronger coupling with the environment and operates at shorter relaxation times that are comparable to the environment correlation time-scales: they are, therefore, inherently non-Markovian, and cannot be described by Lindbladian completely positive semigroup maps \cite{wolf2008,rivas2010,rivas2014}.
Non-Markovianity is an essential feature in the open dynamics of more complex quantum systems, such as photosynthetic pigment protein complexes \cite{chin2013,chen2014},
quantum dots in photonic-crystal cavities \cite{madsen2013},
and other biological \cite{thorwart2009,huelga2013} and many-body strongly correlated systems \cite{rivas2010}.
Further, non-Markovian dynamics allow for an active backflow of information from the environment to the system \cite{breuer2009,lofranco2012,addis2013,haseli2014}, which has significant implications in quantum information science as observed by its role in quantum key distribution \cite{vasile2011b}, metrology \cite{matsuzaki2011,chin2012}, quantum Darwinism \cite{darwin2015}, preservation of quantum correlations \cite{bellomo2007,dijkstra2010}, thermodynamical work extraction \cite{Bylicka2015}, and in designing enhanced quantum protocols \cite{huelga2012,laine2014,bylicka2014}.

One of the primary difficulty in analyzing non-Markovian dynamics is the complicated mathematical description of the dynamical quantum maps, as compared to Markovian processes \cite{rivas2014, vacchini2011, addis2014}. This currently limits the analytical solution of non-Markovian maps to only a few quantum models. However, these models are of immense importance as they offer a realistic characterization for a variety of open systems, with a huge potential in practical applications \cite{vasile2011b,matsuzaki2011,chin2012,darwin2015,bellomo2007,dijkstra2010,huelga2012,laine2014,
bylicka2014, Bernardes2014, Cerrillo2014, Chruscinski2014}. These theoretical investigations have been developed, in parallel, with a substantial increase in engineered environments that allow experimental modelling and control of non-Markovian dynamics \cite{myatt2000,wang2005,frank2009,lucas2013,sweke2014,eisert2015, Bernardes2015, Xiong2015}.

More generally, in recent years, there have been several studies focused on the characterization and possible {\it quantification} of non-Markovianity in open quantum system dynamics \cite{rivas2014,breuer2015}.
The main focus of several works on non-Markovianity has been to capture the deviation from the dynamical semigroup properties of the completely positive trace preserving (CPTP) quantum maps \cite{wolf2008,rivas2010}. In other words, a defining feature of Markovian dynamics is the divisibility of the CPTP map, and any departure from this property is a valid indicator of non-Markovianity \cite{rivas2010}. Hence, the non-monotonic behaviour under non-divisible CPTP maps of a suitable quantum information-theoretic quantity,  such as distinguishability \cite{breuer2009}, entanglement \cite{rivas2010}, or quantum mutual information \cite{luo2012}, can be used to witness and quantify non-Markovianity. Other definitions have exploited the dynamical behavior of Fisher information \cite{lu2010}, accessible information \cite{fanchini2014}, local uncertainty \cite{zhi2014}, and interferometric power \cite{dhar2015} (for other witnesses and quantifiers, see \cite{raja2010, UshaDevi2011, Devi2012, smirne2013}). However, the various measures and witnesses do not reliably embody necessary and sufficient conditions for all non-Markovian processes, and in general may not be compatible with each other \cite{rivas2014, vacchini2011, addis2014,breuer2015}. It is thus desirable to have a more concrete theoretical formalism for non-Markovianity in quantum processes, and further studies are required to fill the gap in the contemporary understanding of the subject.

Though the majority of studies have sought to address the question of non-Markovianity in the dynamics of discrete level open quantum systems, such as qubits and spin chains, an analogous characterization for continuous variable (CV) systems remains much less developed, despite the fact that CV systems, and in particular Gaussian states thereof, constitute fundamental and highly controllable resources for a plethora of quantum information and communication protocols \cite{braun2005,weedbrook2012,Adesso2014ext}. However, in recent times, there have been a few attempts to extend the theoretical formalism to characterize non-Markovianity in discrete level systems to the CV scenario \cite{vasile2011,lorenzo2013,cazzaniga2013,torre2015,eisert2015}. Most notably, approaches to capture the non-monotonicity of distinguishability \cite{vasile2011} and non-divisibility of dynamical maps \cite{torre2015,eisert2015} have been used to define important criteria to witness and measure non-Markovianity of Gaussian quantum dynamical maps, i.e., maps preserving the Gaussianity of their inputs. Attempts have also been made to characterize non-Markovianity in terms of the universality of Gaussian dynamical maps \cite{cazzaniga2013} and volume of the physical Gaussian state space \cite{lorenzo2013}.

In this paper, we seek to define a measure of non-Markovianity for Gaussian quantum channels using an operational figure of merit defined in the context of quantum metrology, namely the Gaussian interferometric power (GIP) \cite{Girolami2014,Adesso2014,Bera2014}. The GIP quantifies the guaranteed precision achieved using bipartite Gaussian probes in a black-box interferometry setting, where the generator of the phase shift to be estimated is \emph{a priori} unknown. The non-monotonic evolution of the interferometric power has been very recently shown,  by some of us, to be useful to construct a valid witness and quantifier of non-Markovianity in discrete level open systems, efficiently computable in the case of single-qubit dynamics \cite{dhar2015}.
In the present work, we extend the formalism to investigate the specific conditions that characterize the parameter regimes related to Markovian and non-Markovian dynamics in Gaussian channels. We consider two paradigmatic CV Gaussian maps based on the damping master equation \cite{ferraro2005} and the quantum Brownian motion \cite{maniscalo2004,eisert2015}. 
We observe that our characterization, based on GIP, allows one to define a computable witness of non-Markovianity for the considered Gaussian dynamics, consistent with the necessary and sufficient conditions for non-Markovian dynamics provided by non-divisibility of the Gaussian maps. We find that non-Markovianity can be efficiently witnessed even with nonentangled probes. Further, we investigate the optimal Gaussian probes that are needed to obtain a quantitative measure of non-Markovianity and analyze the scaling of this quantity with the mean energy of the probes, thus discussing the role of entanglement for enhanced sensitivity in the detection of non-Markovianity.  We also comment on the robustness of our results with varying channels parameters, including the response to finite bath temperatures.

The organization of the paper is as follows. In Sec.~\ref{gauss}, we present a brief account on Gaussian states and dynamical maps, followed by a description of the GIP measure. 
Next, in Sec.~\ref{nonmarkovs}, we present the characterization of non-Markovianity for Gaussian channels and we illustrate the framework of this paper in two examples: the damping master equation with a single decay parameter and the quantum Brownian motion. We conclude with a discussion on the results in Sec.~\ref{concl}.

\section{Gaussian states, Gaussian channels, and Gaussian interferometric power}
\label{gauss}

In this section, we briefly introduce the basic mathematical formalism to describe Gaussian quantum states and dynamical maps that are relevant to CV quantum information and the present investigation. In our work, we study non-Markovianity in CV systems by  focusing on Gaussian states and Gaussian channels and by analyzing the evolution of the GIP.

\subsection{Gaussian states}

A CV system of two modes $A$ and $B$ (with annihilation operators $\hat{a}$ and $\hat{b}$ respectively), can be defined by the quadrature vector $\hat{\boldsymbol{O}} = \{ \hat{q}_A, \hat{p}_A,  \hat{q}_B, \hat{p}_B\}$, where $\hat{q}_k = (\hat{a}_k + \hat{a}_k^\dagger)/\sqrt{2}$ and $\hat{p}_k = (\hat{a}_k - \hat{a}_k^\dagger)/{\sqrt{2}i}$, where $k = A, B$ (assuming natural units, $\hbar = 1$). The quadratures obey the canonical commutation relations $ [\hat{O}_j, \hat{O}_k] = i \Omega_{jk}$, with the two-mode symplectic form
\begin{equation}
\boldsymbol{\Omega} =
\left(
\begin{array}{cc}
0 & 1 \\                                                                                                                                                               -1 & 0 \\                                                                                                                                                              \end{array}                                                                                                                                                            \right)^{\bigoplus  2}.
\end{equation}
A Gaussian state $\rho_{AB}$ \cite{adesso2007, Adesso2014ext, cerf2007, paris2005} is represented by a Gaussian characteristic function in phase space, and is completely characterized by its first and second statistical moments of the quadrature vector, given respectively by the displacement vector $\boldsymbol{\delta}_{AB} = (\delta_j)$ and the covariance matrix  $\boldsymbol{\sigma}_{AB} = (\sigma_{jk})$, where $\delta_j = \tr [\rho_{AB} \hat{O}_j]$ and $\sigma_{jk} = \tr[\rho_{AB} \{ (\hat{O}_j - \delta_j), (\hat{O}_k - \delta_k) \}_+]$ (with $j, k = 1, \ldots,4$ for two modes), and $\{\cdot,\cdot\}_+$ is the anti-commutator. With no loss of generality, one can set the first moments as null and, for all informational purposes, any Gaussian state can be completely determined by its covariance matrix. A \emph{bona fide} condition satisfied by all physical Gaussian states is the Robertson-Schr\"{o}dinger uncertainty relation, given by
\begin{equation}
\boldsymbol{\sigma}_{AB} + i \boldsymbol \Omega \geq 0.
\end{equation}
Its worth recalling that, by local symplectic operations (equivalent to local changes of basis on the state), every two-mode covariance matrix can be transformed to a standard form, with diagonal $2 \times 2$ subblocks that can be written as:
\begin{equation}
\boldsymbol{\sigma}_{A B} = \left(                                                                                                                                                              \begin{array}{cc}                                                                                                                                                               \boldsymbol{\alpha} & \boldsymbol{\gamma} \\                                                                                                                                                                \boldsymbol{\gamma}^T & \boldsymbol{\beta} \\                                                                                                                                                              \end{array}                                                                                                                                                            \right),
\label{CMgeneral}\end{equation}
where $\alpha = \textrm{diag} \{a,a\}$, $\beta = \textrm{diag} \{b,b\}$, $\gamma = \textrm{diag} \{c,d\}$, such that $a,b \geq 1$, $c \geq \abs{d} \geq 0.$ The total mean number of excitations (proportional to the total mean energy) of a two-mode Gaussian state can be defined as: $E \equiv \bar{n}_A + \bar{n}_B = 2 \bar{n},$ where $\bar{n}_A = (\tr [\boldsymbol{\alpha}] - 2)/4$ and $\bar{n}_B = (\tr [\boldsymbol{\beta}] - 2)/4$ are the mean number of excitations in modes $A$ and $B$ respectively, and $\bar{n}$ denotes the mean number of excitations per mode. Throughout this paper we will always impose the physical assumption that any initial state be constrained to a  \emph{finite} mean energy.

\subsection{Gaussian channels}

In this work we study the open dynamics of Gaussian states through Gaussian channels, i.e. quantum channels that map Gaussian states into Gaussian states. If the final evolved state $\rho_{AB}(t)$ is Gaussian, its properties can similarly be studied through its covariance matrix. The dynamical  evolution of a two-mode covariance matrix $\boldsymbol{\sigma}$ subjected to a Gaussian channel can be written as \cite{torre2015,Lindblad2002,heinosaari2010}:
\begin{equation}
\boldsymbol{\sigma}(0) \mapsto \boldsymbol{\sigma}(t)  = \boldsymbol{X}(t)~ \boldsymbol{\sigma}(0) \boldsymbol{X}(t)^T + \boldsymbol{Y}(t),
\label{xmap}
\end{equation}
where $\boldsymbol{X}, \boldsymbol{Y}$ are real $4 \times 4$ matrices (for $n$-mode systems, $2n \times 2n$ matrices). For completely positive maps, the matrices must satisfy \cite{torre2015,Lindblad2002,heinosaari2010}:
\be
\boldsymbol{Y} + i \boldsymbol\Omega - i \boldsymbol X \boldsymbol\Omega \boldsymbol{X}^T \geq 0.
\label{xcond}
\ee
If, as in the case considered in this paper, only mode $A$ is subjected to a Gaussian channel, the evolved two-mode state is mapped to
\begin{eqnarray}
\boldsymbol{\sigma}_{AB}(t) &=& [(\sqrt{\Lambda_1(t)} \mathbb{I}_A) \oplus \mathbb{I}_B]^T  \boldsymbol{\sigma}_{AB}(0) [(\sqrt{\Lambda_1(t)} \mathbb{I}_A) \oplus \mathbb{I}_B]\nonumber\\
&+& \Lambda_2(t) \mathbb{I}_A \oplus \mathbb{O}_B,
\end{eqnarray}
where $\boldsymbol{\sigma}_{AB}(0)$ is in the standard form Eq.~(\ref{CMgeneral}), $\mathbb{I}_A$ is the $2 \times 2$ identity matrix acting on mode $A$, $\mathbb{O}_B$ is a null $2 \times 2$ matrix acting on mode $B$, and finally $\Lambda_1(t)$ and $\Lambda_2(t)$ are terms related to the specific dynamical evolution. Explicitly, one can write the evolved covariance matrix as follows,
\begin{equation}
\boldsymbol{\sigma}_{A B}(t) = \left(
\begin{array}{cccc}
a\Lambda_1(t)+\Lambda_2(t)&\hspace{-0.2cm}0&\hspace{-0.2cm}c\sqrt{\Lambda_1(t)}&\hspace{-0.2cm}0\\
0&\hspace{-0.2cm}a \Lambda_1(t)+\Lambda_2(t)&\hspace{-0.2cm}0&\hspace{-0.2cm}d\sqrt{\Lambda_1(t)}\\
c\sqrt{\Lambda_1(t)}&\hspace{-0.2cm}0&\hspace{-0.2cm}b&\hspace{-0.2cm}0\\
0&\hspace{-0.2cm}d\sqrt{\Lambda_1(t)}&\hspace{-0.2cm}0&\hspace{-0.2cm}b
\end{array}
\right).
\label{mapsigma}
\end{equation}

\subsection{Gaussian Interferometric Power}
\label{Gip}

The GIP quantifies the ability of a two-mode Gaussian probe to estimate a local phase shift in a worst case scenario \cite{Adesso2014,Bera2014}, according to an operational setting generally known as black-box interferometry \cite{Girolami2014}. In the CV Gaussian scenario, the protocol to define the GIP can be summarized as follows: a two-mode Gaussian state $\rho_{AB}$ is prepared by two parties Alice and Bob as a probe for an interferometer; mode $B$ (which we call the ancilla) enters a black-box in which it undergoes a unitary transformation $\hat{U}_B^\phi = e^{-i \phi \hat{H}_B}$, where the parameter $\phi$ is unknown, and only the spectrum of the generator $\hat{H}_B$ (assumed nondegenerate to avoid trivial dynamics) is initially known. Fixing the spectrum of $\hat{H}_B$ to be harmonic, and restricting to Gaussianity preserving transformations, the unitary $\hat{U}_B^\phi$ can be written as $\hat{U}_B^\phi = \hat{V}_B^\dagger \hat{W}_B^\phi  \hat{V}_B$, where $\hat{W}_B^\phi = e^{-i \phi  \hat{b}^\dagger \hat{b}}$ is a conventional phase shift operator and $\hat{V}_B$ is an arbitrary Gaussian unitary transformation. The transformed two-mode state is given by
\begin{equation}
\rho_{AB}^{\phi,\hat{V}_B} = (\mathbb{I}_A \otimes \hat{U}_B^{\phi})~ \rho_{AB}~ (\mathbb{I}_A \otimes \hat{U}_B^{\phi})^\dagger.
\label{rhotransformed}
\end{equation}
The information on the black-box generator is provided to the two parties only after the transformation (i.e., the choice of $\hat{V}_B$ is disclosed), thus allowing for optimal measurements to be performed on the output. For any given setting of $\hat{V}_B$, the objective of the interferometric setup is to deduce the unknown phase $\phi$ with the maximum possible precision, i.e., to construct the best possible estimator of the parameter. Assuming a large number $\kappa$ of copies of the probing state $\rho_{AB}$ are initially prepared, and letting the interferometric trial be repeated accordingly $\kappa$ times, by collective processing one can construct an estimator  $\phi_{est}$ whose variance $\Delta \phi^2 =  \langle (\phi_{est} - \phi)^2 \rangle$ is constrained by the Cram\'{e}r-Rao bound,
\begin{equation}\label{cramer}
\kappa \Delta \phi^2 \geq \frac{1}{\mathcal{F}(\rho_{AB}^{\phi,\hat{V}_B})},
\end{equation}
where $\mathcal{F}(\rho_{AB}^{\phi,\hat{V}_B})$ is the quantum Fisher information, which can be defined as
\begin{equation}
\mathcal{F}(\rho_{AB}^\phi) = -2 \lim_{d \phi \rightarrow 0} \frac{\partial^2 F(\rho_{AB}^{\phi,\hat{V}_B}, \rho_{AB}^{\phi+ d \phi,\hat{V}_B})}{\partial (d \phi)^2},
\end{equation}
with $F(\rho_{AB}^\phi, \rho_{AB}^{\phi+d \phi})$ being the Uhlmann fidelity. Since the inequality in (\ref{cramer}) is asymptotically attainable, the quantum Fisher information directly quantifies the precision in the estimation of $\phi$ for each given choice of the black-box generator, being inversely related to the minimum variance of the optimal estimator, and being by construction geometrically interpreted as the rate (speed) at which the probing state changes following an infinitesimal change of the parameter to be estimated.

In the black-box paradigm, the natural figure of merit one adopts to quantify the guaranteed precision allowed by a given probing state $\rho_{AB}$, in the estimation of a parameter $\phi$ with incomplete prior information on the local generator is therefore given by a worst-case scenario, whereby one minimizes the quantum Fisher information over all the local generators (with the fixed spectrum). The resulting quantity is known as the interferometric power of $\rho_{AB}$ \cite{Girolami2014}. In the CV Gaussian setting, of relevance in this paper, the GIP of the two-mode Gaussian state $\rho_{AB}$, with respect to mode $B$, is thus defined as:
\begin{equation}\label{gippone}
\mathcal{Q}_{B}^{G}({\rho_{AB}}) = \frac{1}{4} \inf_{\hat{V}_B} \mathcal{F} (\rho_{AB}^{\phi, \hat{V}_B}).
\end{equation}
It has been proven that, both in finite-dimensional and in CV systems, the interferometric power is a measure of discord-type correlations for general mixed states $\rho_{AB}$, reducing to a measure of entanglement in the particular case of pure states \cite{Girolami2014,Adesso2014,Bera2014}. In the Gaussian case, the GIP $\mathcal{Q}_{B}^G$ satisfies in particular the following properties \cite{Adesso2014,Bera2014}: (i) it vanishes iff $\rho_{AB}$ is a product state (since the only classically correlated Gaussian states are product states \cite{Adesso2010,Mista2014}); (ii) it is invariant under local unitaries; (iii) it is monotonically nonincreasing under local CPTP operations on party $A$. The latter property is the crucial one which will allow us to construct a non-Markovianity indicator based on the GIP in the next Section.

A handy feature of the GIP is that the minimization in Eq.~(\ref{gippone}) can be solved exactly for two-mode Gaussian states and local Gaussian unitaries. The result is a compact formula to calculate the GIP in terms of the covariance matrix $\boldsymbol{\sigma}_{AB}$ of an arbitrary two-mode Gaussian state, given by \cite{Adesso2014}
\begin{equation}
\mathcal{Q}_B^G (\boldsymbol{\sigma}_{A B}) = \frac{C_x + \sqrt{C_x^2 + C_y C_z}}{2 C_y},
\label{GIP}
\end{equation}
where
\begin{eqnarray}
C_x &=& (I_2 + I_3) (1 + I_1 + I_3 - I_4) - I_4^2 \nonumber \\
C_y &=& (I_4-1) (1 + I_1 + I_2 + 2 I_3 + I_4) \nonumber \\
C_z &=& (I_2+I_4) (I_1 I_2 - I_4) + I_3 (1 + I_1) (2 I_2 + I_3) ,
\end{eqnarray}
with the symplectic invariants of the general covariance matrix \eqref{CMgeneral} given by $I_1 = \det \boldsymbol{\alpha}$, $I_2 = \det \boldsymbol{\beta}$, $I_3 = \det \boldsymbol{\gamma}$, and $I_4 = \det \boldsymbol{\sigma}_{AB}$.

%

\section{Characterization of Non-Markovianity in Gaussian channels}\label{nonmarkovs}

\subsection{Non-divisibility of Gaussian channels}

In this section, we focus on the non-Markovian characteristics of a Gaussian quantum channel in the open system dynamics of CV states. The time evolution of an initial Gaussian state $\rho(0)$, under a Gaussian dynamical map $\Lambda_t$, is given by the relation $\rho(t) = \Lambda_t \rho(0)$, where $\Lambda_t$ is CPTP. Under Markovian dynamics, $\Lambda_t$ can be described by Lindbladian dynamical semigroups and the quantum map is divisible. Hence, a Markovian channel can be written as $\Lambda_t = \Lambda_{t,t'} \Lambda_{t',0}$, for $t > t' > 0$  \cite{addis2014, rivas2014}, which implies that the CPTP map describing the dynamics of the system from time 0 to $t$ can be divided into two CPTP maps from initial time 0 to some intermediate time $t'$, and from $t'$ to the final time $t$. This implies that  any quantum informational quantity that is by definition contractive under CPTP maps, must be strictly monotonic under a Markovian evolution.

Hence, either by observing a violation of the divisibility criterion, or by witnessing a breakdown of monotonicity of a contractive quantum information measure, one can get natural witnesses of non-Markovianity; a few proposals along these lines have been reported recently for CV systems \cite{vasile2011,lorenzo2013,cazzaniga2013,torre2015,eisert2015}.
In our work, we use the non-monotonic evolution of the GIP  to detect and quantify Gaussian non-Markovianity. Before introducing our non-Markovianity indicator, we briefly recall a characterization of non-divisible Gaussian channels.


Given a Gaussian dynamical map (see Sec.~\ref{gauss}), acting on the covariance matrix of a  Gaussian state as described in Eq.~\eqref{xmap}, and defined by the matrices $\boldsymbol X$ and $\boldsymbol Y$ respecting the condition of Eq.~(\ref{xcond}), it can be shown that a necessary and sufficient condition for non-Markovianity, due to a violation of the divisibility criterion \cite{rivas2010}, is given by  \cite{torre2015}
\begin{equation}
\boldsymbol Y(t+\epsilon) + i \boldsymbol \Omega - i \boldsymbol X(t+\epsilon)~ \boldsymbol \Omega~ \boldsymbol X^T(t+\epsilon) < 0,
\label{c-div}
\end{equation}
where $\epsilon$ is any intermediate time instant and $\boldsymbol\Omega$ is the symplectic matrix defined earlier. While Eq.~(\ref{c-div}) is useful to detect a non-Markovian evolution, the authors of Ref.~\cite{torre2015} have further proposed to  {\it quantify} the degree of non-Markovianity in CV Gaussian channels by a measure defined as
\be
\mathcal{N}_{D} = \int_I G(t) dt,
\label{divisibility}
\ee
where the integral is performed over the time interval $I$, and $G(t) = \sum_{k} \frac{1}{2} \lim_{\epsilon \rightarrow 0^+} [\abs{\nu_k (t + \epsilon, t)} - \nu_k (t + \epsilon, t)]$, with $\nu_k (t + \epsilon, t)$ being the eigenvalues of the symmetric matrix in the left hand side of Eq.~\eqref{c-div}.

\begin{figure}[t]
  \centering
  \includegraphics[width=8.5cm]{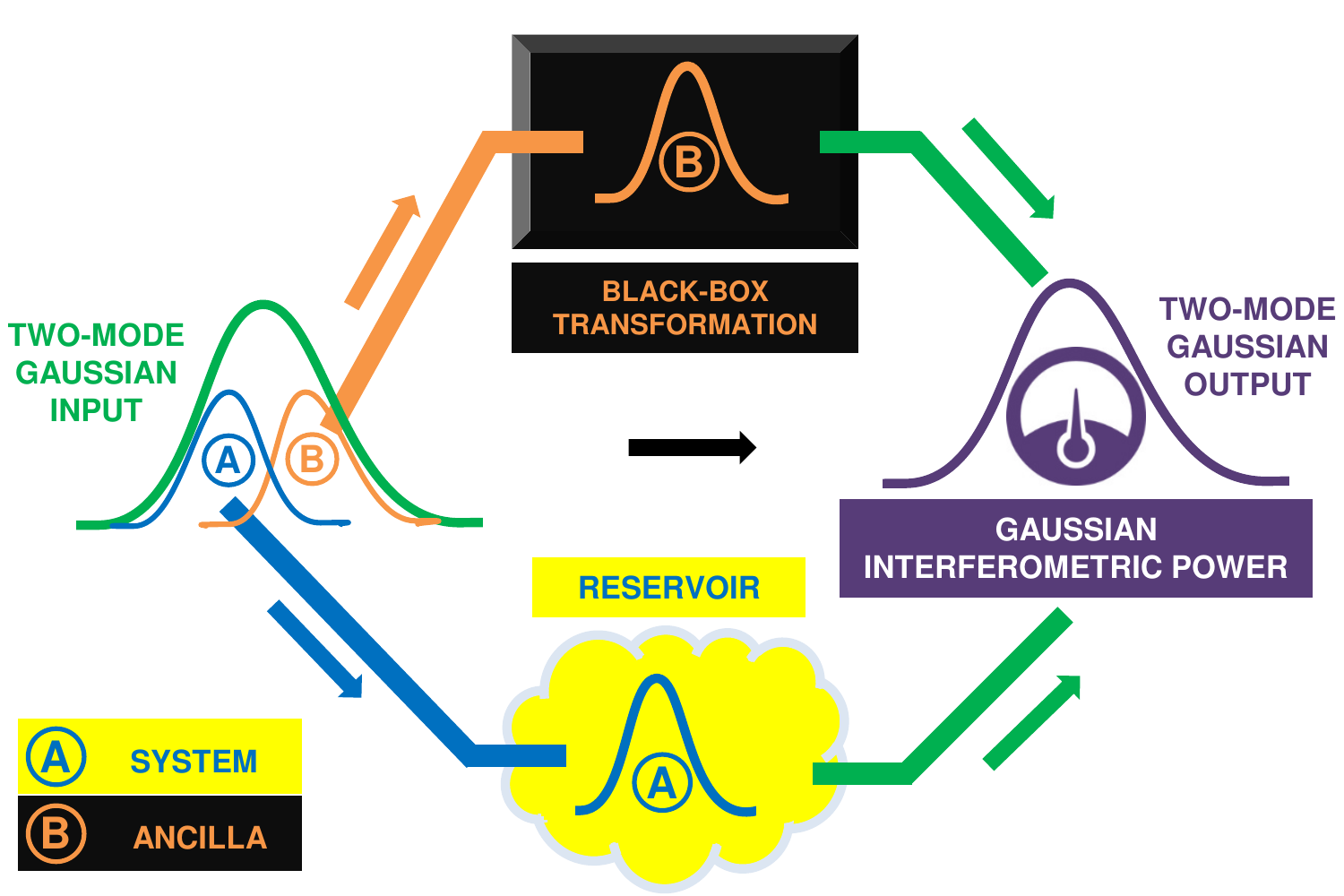}
  \caption{(Color online) A schematic illustrating the introduced protocol to characterize continuous variable non-Markovianity via the Gaussian interferometric power of a two-mode Gaussian state. The system mode $A$ is subjected to a local quantum Gaussian dynamical map, modelling the interaction with the reservoir, whereas the ancillary mode $B$ is subjected to an unknown Gaussian unitary transformation. The observation of a non-monotonic time evolution of the  Gaussian interferometric power of the output two-mode state yields a valid witness of non-Markovian dynamics on the system mode.}\label{setup}
\end{figure}

\subsection{Witnessing and quantifying non-Markovianity via the Gaussian interferometric power}

We define an alternative measure of non-Markovianity in CV systems by making use of the non-monotonic behavior of the GIP under nondivisible CPTP Gaussian maps. The interferometric power has been shown to be a practical quantifier of non-Markovianity in the context of discrete variable systems \cite{dhar2015}. Here we follow a similar approach to investigate non-Markovianity in Gaussian CV systems; see Fig.~\ref{setup} for an illustration of the setup. Let us consider a Gaussian state $\rho_{AB}$ of the two modes $A$ and $B$, where the system mode $A$ undergoes an open evolution (due to the presence of a reservoir) modelled as a Gaussian CPTP map,  while the ancillary mode $B$ is sent through a black-box unitary. As described in Section~\ref{Gip}, the GIP for the output two-mode state $\rho_{AB}(t)$ can be exactly computed; this quantity captures the ability of such  two-mode state to serve as a probe for the estimation of a phase shift imprinted during the black-box transformation on the ancilla mode, in a worst-case scenario. Due to its contractivity under local channels on the system mode $A$, the GIP  $\mathcal{Q}_B^G (\rho_{AB}(t))$ is by definition a monotonically nonincreasing function of time under any local divisible CPTP map acting on mode $A$. Hence, under Markovian Gaussian channels, the time-derivative of the GIP is strictly non-positive, i.e.,
$\frac{d}{dt} \mathcal{Q}^G (\rho_{AB}(t)) \le 0$, for all $t \ge 0$.
The non-Markovian character in the dynamics of mode $A$ is thus characterized by the violation of the monotonic behavior of the GIP. If one defines the quantity
\begin{equation}
\mathcal{D}(t) = \frac{d}{dt} \mathcal{Q}_B^G (\rho_{AB}(t)),
\label{flow}
\end{equation}
any positive value of  $\mathcal{D}(t)$ captures the non-monotonic behavior of the GIP under the considered Gaussian channel. Hence, for a given initial Gaussian state $\rho_{AB}(0)$ with a covariance matrix $\boldsymbol\sigma_{AB}(0)$, and denoting the evolved state as  $\rho_{AB}(t) = (\Lambda_A \otimes \mathbb{I}_B) \rho_{AB}(0)$ following the action of a local Gaussian channel $\Lambda_A$ on the system mode $A$, we define a metrological {\it witness} of Gaussian non-Markovianity by the quantity
\be
\mathcal{N}_Q^{\boldsymbol{\sigma}} (\Lambda) = \int_{\mathcal{D}(t)>0} \mathcal{D}(t) dt,
\label{witness}
\ee
where the integral is over all the time instants $t$ such that  $\mathcal{D}(t)>0$. The superscript $\boldsymbol{\sigma}$ in $\mathcal{N}_Q^{\boldsymbol{\sigma}}$ emphasizes the fact that the witness is specific to a given initial Gaussian state  with a covariance matrix $\boldsymbol{\sigma} \equiv \boldsymbol\sigma_{AB}(0)$. Following \cite{breuer2009,rivas2010}, the non-Markovianity witness in Eq.~(\ref{witness}) can be optimized over the set of all initial Gaussian states to obtain a proper  {\it measure} of Gaussian non-Markovianity, defined as
\begin{equation}
\mathcal{N}_Q (\Lambda) = ~\max_{\boldsymbol\sigma_{AB}(0)} \mathcal{N}_Q^{\boldsymbol\sigma} (\Lambda)~.
\label{nonmarkov}
\end{equation}
Numerically, the integral may be evaluated in small interval integrals:
\begin{equation}
\mathcal{N}_Q(\Lambda) = \max_{\boldsymbol\sigma_{AB}(0)} \sum_{k; \mathcal{D}(t) > 0} \int_{t_i^k}^{t_f^k} \mathcal{D}_k (t) dt.
\end{equation}


In the following subsections we show how the above witness and measure of CV non-Markovianity can be applied to investigate the characteristics of physical quantum maps. For the purposes of our investigation, we consider two fundamental Gaussian channels that are based on the damping master equation  \cite{ferraro2005} and the quantum Brownian motion  \cite{maniscalo2004,eisert2015}.

\subsection{Non-Markovianity in the damping model with a single decay parameter}

We begin with a Gaussian channel characterized by a single time-dependent decay parameter, which is represented by a well-described damping master equation, given by
\be
\frac{d \rho}{dt} = \alpha \frac{\gamma(t)}{2} (2 \hat{a} \rho \hat{a}^\dagger - \{\hat{a}^\dagger \hat{a}, \rho\}_+), \label{DME}
\ee
where $\alpha$ is the coupling constant, and $\gamma(t)$ the time-dependent damping coefficient. It can be easily shown that for the damping master equation the divisibility property is satisfied by the condition $\gamma(t) \geq 0,~  \forall~ t \geq 0$ \cite{vasile2011, torre2015}. Thus, the damping master equation provides an excellent model to test the validity of any prospective non-Markovian quantifier.

\begin{figure}
  \centering
  \includegraphics[width=7.5cm]{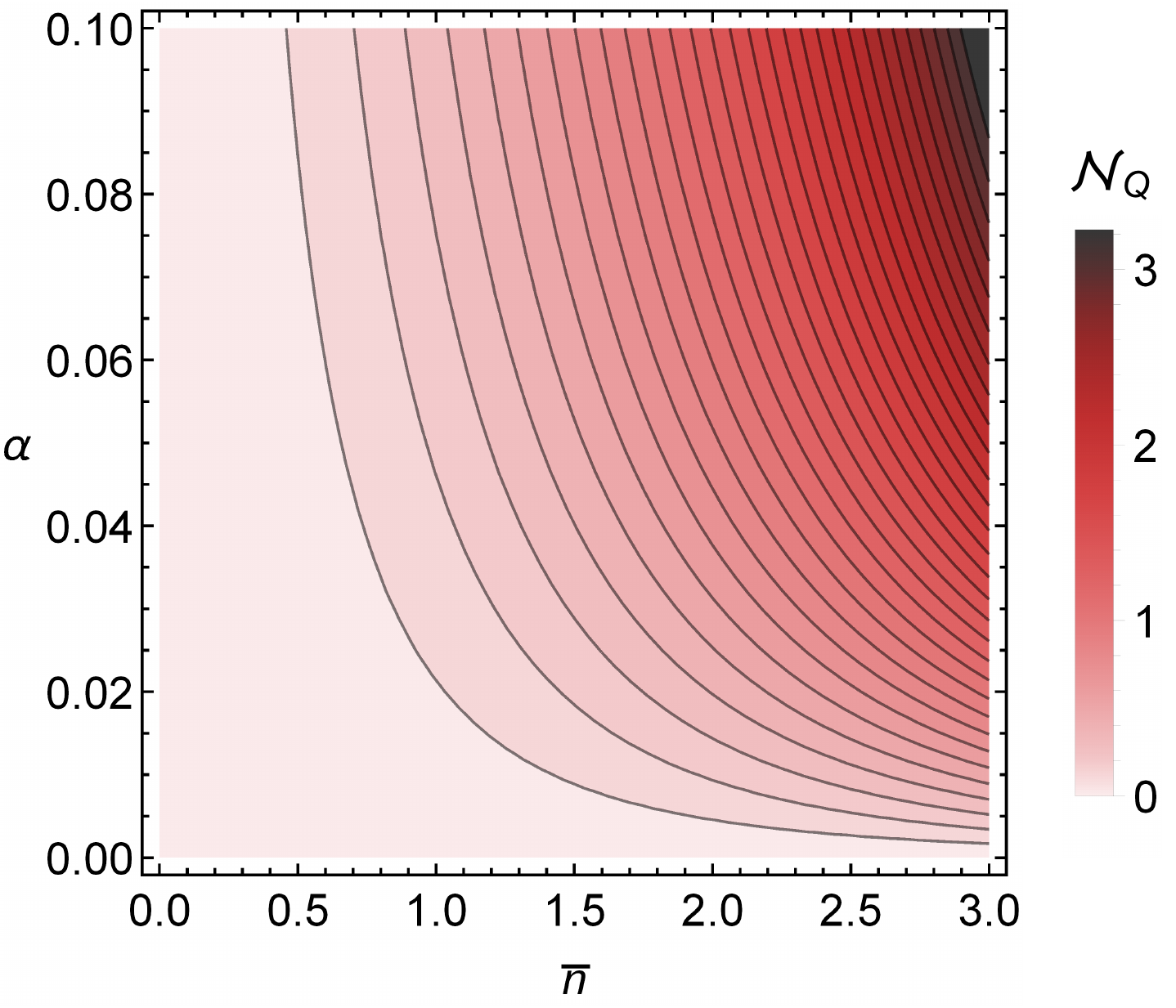}\\
  \caption{(Color online) Measure of non-Markovianity $\mathcal{N}_Q$ for the damping master equation, as a function of the mean number of excitations $\bar{n}$ and of the coupling constant $\alpha$. All the quantities plotted are dimensionless.}\label{damping1}
\end{figure}

In our setting, the covariance matrix of a two-mode Gaussian state whose mode $A$ is subject to the damping master equation of Eq.~\eqref{DME}, is mapped as: \bq \boldsymbol\sigma_{AB}(t) &=& [(\ei^{-x(t)/2} \mathbb{I}_A) \oplus \mathbb{I}_B]^T \boldsymbol\sigma_{AB}(0) [(\ei^{-x(t)/2} \mathbb{I}_A) \oplus \mathbb{I}_B] \nonumber \\ && + (1 - \ei^{-x(t)}) \mathbb{I}_A \oplus \mathbb{O}_B, \label{mapsigma1}\eq where \be x(t) = \alpha \int_0^t 2 \gamma(s) ds. \ee
If the initial covariance matrix $\boldsymbol\sigma_{AB}(0)$ is in standard form, as in Eq.~\eqref{CMgeneral}, then after local  damping the covariance matrix is mapped by Eq.~\eqref{mapsigma1} to the evolved state $\sigma_{A B}(t)$ given explicitly by
\be
\boldsymbol{\sigma}_{A B}(t) = \left(
\begin{array}{cccc}
a(t) & 0 & c(t) & 0 \\
0 & a(t) & 0 & d(t) \\
c(t) & 0 & b & 0 \\
0 & d(t) & 0 & b \\
\end{array}
\right),
\ee
where $a(t) = a \ei^{-x(t)} + (1 - \ei^{-x(t)})$, $c(t) = c \ei^{-x(t)/2}$, $d(t) = d \ei^{-x(t)/2}$. To illustrate quantitatively our characterization of non-Markovianity as measured using the GIP,  we can choose on  a specific instance of  damping regime, given by \cite{vasile2011}
\be \gamma(t) = \frac{1}{2} \times \Bigg\{
\begin{array}{ll}
\ei^{-t/10} \sin t, & \quad \text{if $t < 5 \pi /2$}\\
\ei^{-\pi/4}, & \quad \text{if $t \geq 5 \pi/2$}.\\
\end{array} \ee
In this case, there exists only a single interval, $\pi < t < 2 \pi$, where $\gamma(t)$ is negative. Therefore, the non-Markovian witness $\mathcal{N}_Q^{\boldsymbol\sigma}$ and measure $\mathcal{N}_Q$, based on GIP, can be analytically evaluated by the expressions
\begin{eqnarray}
\mathcal{N}_Q^{\boldsymbol\sigma} (\Lambda)&=& \mathcal{Q}_B^G({\boldsymbol\sigma}_{AB}(t = 2 \pi)) - \mathcal{Q}_B^G({\boldsymbol\sigma}_{AB}(t = \pi)), \label{witnessDME} \\
\mathcal{N}_Q(\Lambda) &=& \max_{{\boldsymbol\sigma}_{AB}(0)} \mathcal{N}_Q^{\boldsymbol\sigma}. \label{dampingNM}
\end{eqnarray}

Figure~\ref{damping1} shows the measure of non-Markovianity, $\mathcal{N}_Q$, for the damping master equation as a function of the mean number of excitations per mode, $\bar{n}$, and of the coupling constant $\alpha$. We note that, for fixed $\alpha$, the value of $\mathcal{N}_Q$ increases with higher values of the mean energy. Similarly, for a constant mean energy $\bar{n}$, $\mathcal{N}_Q$ increases monotonically with $\alpha$.

In the next subsection we provide more details on the optimal input states to witness non-Markovianity, and we carefully discuss the role of entanglement in the detection.
%
%

\subsection{\label{beyent}Non-Markovianity witness beyond entangled probes}

\begin{figure}
  \centering
  \includegraphics[width=8.5cm]{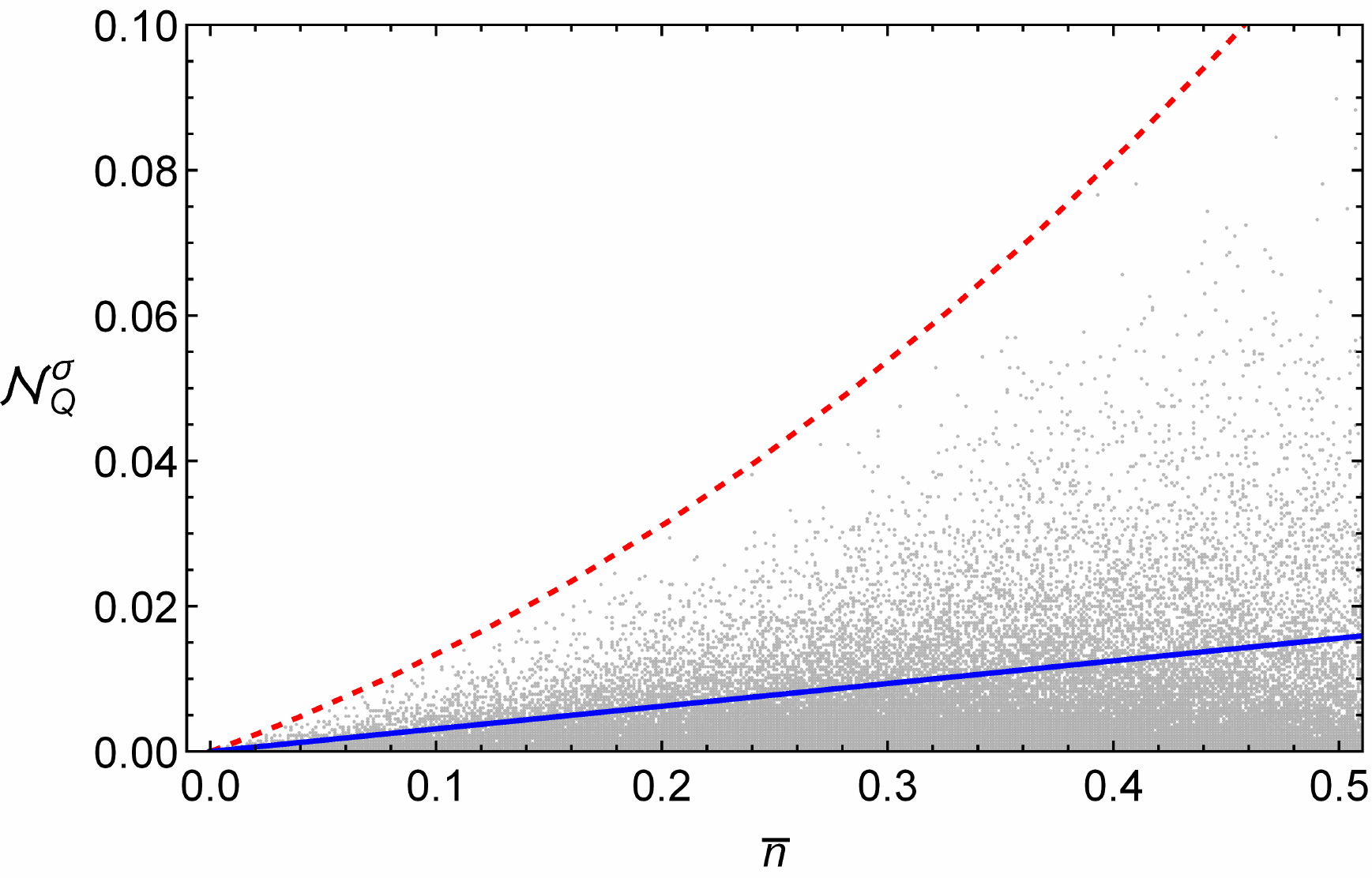}\\
  \caption{(Color online) Witness of non-Markovianity for $10^5$ randomly generated two-mode Gaussian states, $\sigma_{AB}(0)$, captured by $\mathcal{N}_Q^{\boldsymbol\sigma}$ as defined in Eq.~\eqref{witnessDME}, for the damping master equation. $\mathcal{N}_Q^{\boldsymbol\sigma}$ is plotted as a function of the mean number of excitations $\bar{n}$ for a fixed $\alpha = 0.1$. The (red online) dashed line represents the pure two-mode squeezed states and the (blue online) solid line is for the two-mode mixed thermal states. It is evident from the figure that pure two-mode squeezed states optimize the witness $\mathcal{N}_Q^{\boldsymbol\sigma}$, to realize the non-Markovianity measure $\mathcal{N}_Q$ plotted in Fig.~\ref{damping1}. All the quantities plotted are dimensionless.}\label{damping2}
\end{figure}

A remarkable aspect of characterizing non-Markovianity using GIP is its ability to witness the non-Markovian dynamics of a local Gaussian channel by using two-mode probes which exhibit quantum correlations beyond entanglement \cite{Adesso2014}. To illustrate this, we can focus on two important classes of two-mode Gaussian states: the mixed thermal states (MTS), which are always unentangled, and the squeezed thermal states (STS), which can be entangled or separable depending on the trade-off between squeezing and noise.  Both these classes of states can be easily engineered with the current toolbox of quantum optics, using e.g. single-mode squeezing and and linear optical operations, such as beam-splitters \cite{Adesso2014ext,francamentemeneinfischio}. In particular, the MTS can be generated by letting two single-mode thermal states, at different equilibrium temperatures, interact via a balanced 50:50 beam-splitter. The STS can be instead engineered by first squeezing two single-mode thermal states in complementary quadratures, and then letting them interact via a 50:50 beam-splitter. The covariance matrices of these two classes of states can  be explicitly written as,
\begin{eqnarray}
&\boldsymbol{\sigma}_{A B}^{MTS}& = \left(
\begin{array}{cccc}
k_1 e^{2 r_1} x & 0 & k_1 e^{2 r_1} y & 0 \\
0 & k_1 e^{2 r_1} x & 0 & k_1 e^{2 r_1} y \\
k_1 e^{2 r_1} y & 0 & k_1 e^{2 r_1} x & 0 \\
0 & k_1 e^{2 r_1} y & 0 & k_1 e^{2 r_1} x
\end{array}\label{sigmamixed}
\right), ~\textrm{and}
\end{eqnarray}
\begin{eqnarray}
&\boldsymbol{\sigma}_{A B}^{STS}& = \left(
\begin{array}{cccc}
k_2~x' & 0 & k_2~y' & 0 \\
0 & k_2~x' & 0 & - k_2~y' \\
k_2~y' & 0 & k_2~x' & 0 \\
0 & - k_2~y' & 0 & k_2~x'
\end{array}\label{sigmasqueezed}
\right),
\end{eqnarray}
for MTS and STS, respectively, where $x=\cosh (2 r_1)$, $y=\sinh (2 r_1)$, $x'=\cosh (2 r_2)$, and $y'=\sinh (2 r_2)$.
The parameters in Eqs.~\eqref{sigmamixed} and \eqref{sigmasqueezed} are related to the mixedness of the initial thermal states ($k_1, k_2 \geq 1$) and the strength of the Gaussian operations ($r_1, r_2 \geq 0$) such as squeezing (in the STS case) and beam-splitter interaction. Although the MTS states of Eq.~\eqref{sigmamixed} are always separable, and the STS of Eq.~\eqref{sigmasqueezed} are separable when $k_2(x'-y') \geq 1$, both classes of states always contain more general forms of quantum correlations such as quantum discord and have in general a nonzero GIP \cite{Giorda2010,Giorda2012, Adesso2014, Bera2014}, as soon as they are not completely uncorrelated (i.e., as soon as $y, y' \neq 0$). Since the resource ensuring a precision in black-box interferometry is indeed associated with discord-type correlations, as quantified by the GIP, and not with entanglement, and since we are exploiting this figure of merit to investigate non-Markovianity, both classes of states (as well as any other non-product Gaussian state) are in principle able to provide a useful witness of non-Markovianity in generic local Gaussian channels.  There is furthermore a computational advantage in focusing on the two considered classes of states, since for both of them the covariance matrix is in standard form, Eq.~(\ref{CMgeneral}), with $d = \pm c$; in this case, the expression (\ref{GIP}) for the GIP reduces to the much simplified form \cite{Bera2014,Adesso2014}
\begin{equation}
\mathcal{Q}_B^G(\boldsymbol{\sigma}_{AB})|~_{d=\mp c} = \frac{c^2}{2 (a b - c^2 \pm 1)}\,.
\end{equation}
In the specific case of the two classes of states defined above, we have $a = b = k_1 e^{2 r_1} \cosh (2 r_1)$ and $c = k_1 e^{2 r_1} \sinh(2 r_1)$ for MTS, and $a = b = k_2 \cosh (2 r_2), c = k_2 \sinh(2 r_2)$ for STS. As mentioned in Sec.~\ref{gauss}, to issue a fair comparison and to find an appropriate balance between the choice of parameters between these two classes of Gaussian states, we will require the physical assumption that all states are parametrized in terms of finite initial mean energy $E = 2\bar{n}$.

\begin{figure}[h]
  \centering
  \includegraphics[width=8.3cm]{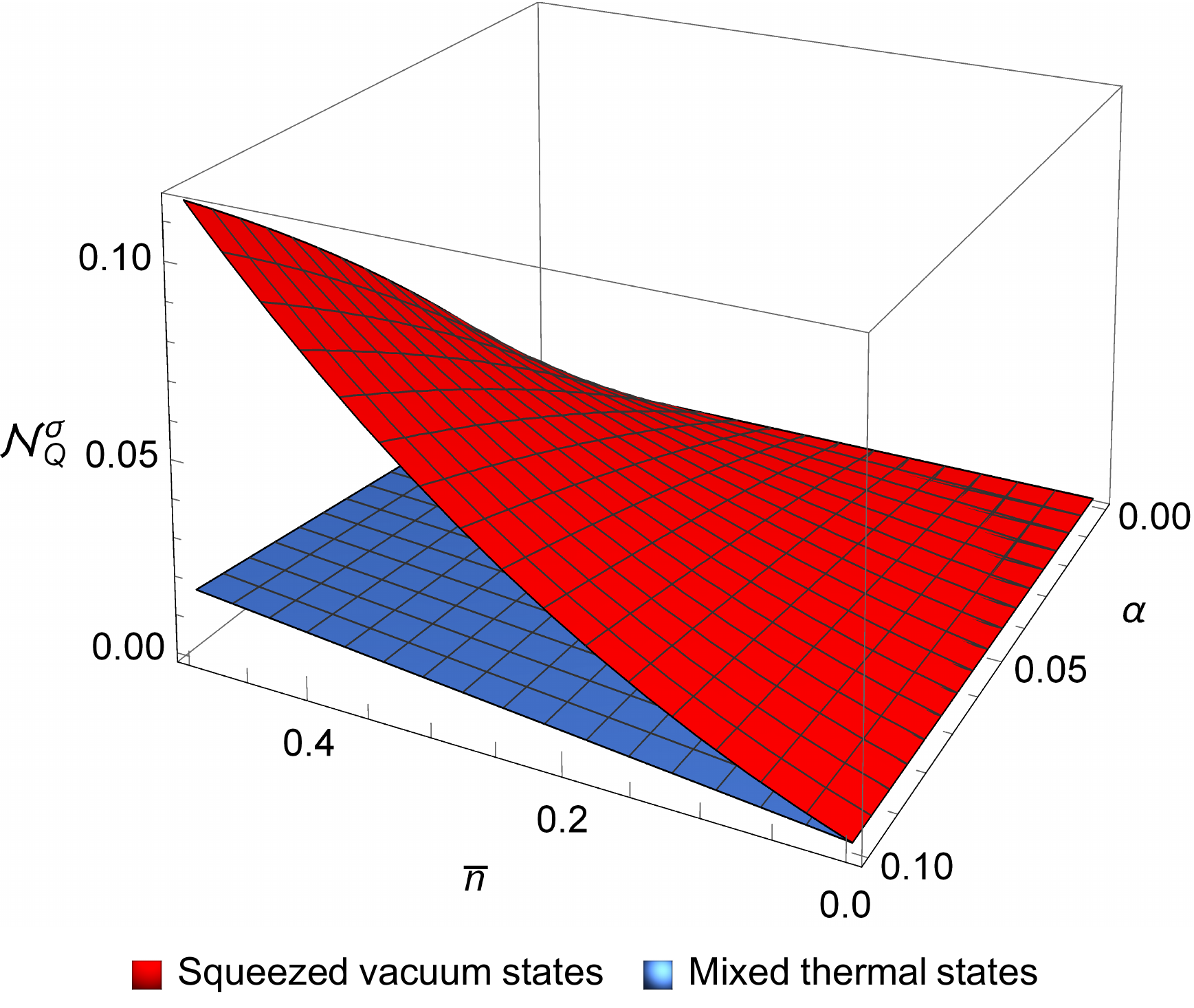}\\
  \caption{(Color online) Witness of non-Markovianity given by $\mathcal{N}_Q^{\boldsymbol\sigma}$ for the damping master equation, as a function of the mean number of excitations $\bar{n}$ and the coupling constant $\alpha$, for the classes of separable two-mode mixed thermal states (blue online lowermost surface)  and entangled two-mode squeezed states (red online uppermost surface), given respectively by
Eqs.~\eqref{sigmamixed} and \eqref{sigmasqueezed}, with $k_1=k_2=1$. All the quantities plotted are dimensionless.}\label{damping3}
\end{figure}

For the damping master equation, we first observe that, within the classes of STS and MTS chosen as initial probe states, the non-Markovianity witness $\mathcal{N}_Q^{\boldsymbol\sigma}$ is analytically optimized (for any fixed $\bar n$), when the states are as pure as possible, i.e., for $k_1=k_2=1$. In this limit, the STS reduce to pure entangled two-mode squeezed vacuum states (also known as twin-beams), with all the energy invested in squeezing, while the separable MTS maintain a nonzero degree of mixedness. We have performed an extensive numerical investigation of the witness $\mathcal{N}_Q^{\boldsymbol\sigma}$ over completely general randomly generated two-mode Gaussian input covariance matrices, in order to identify the globally optimal probes. From the numerical analysis presented in Fig.~\ref{damping2} (in the instance of $\alpha=0.1$), corroborated by the analytical optimization within the classes of STS and MTS states, we conclude that two-mode pure STS (i.e., squeezed vacuum states), spanning the (red online) dashed curve, are the globally optimal probes which attain the maximum in Eq.~(\ref{nonmarkov}), thus achieving the quantification of non-Markovianity of the damping channel through the measure $\mathcal{N}_Q$. This has been verified numerically over an extended range of values of $\alpha$ and $\bar{n}$, and the result obtained by using squeezed vacuum probes matches exactly with the evaluation of the measure in Fig.~\ref{damping1}. Remarkably, though, we can see that the separable MTS states with $k_2=1$, spanning the blue solid curve in Fig.~\ref{damping2}, provide a suboptimal but still significantly nonzero witness of non-Markovianity for the damping master equation. It is interesting to observe that several randomly generated states, even entangled, were found to perform worse than the MTS probes, resulting in significantly lower values of the witness $\mathcal{N}_Q^{\boldsymbol\sigma}$. This shows quite importantly that a robust non-Markovianity witness for Gaussian channels can be generated (at least for the single decay damping model) using separable mixtures of thermal states, which are significantly more economical to prepare than entangled states requiring high degrees of squeezing.

One can then wonder, if entanglement is not needed to detect non-Markovianity of Gaussian channels in our setup, what potential benefit it may bring other than a quantitative enhancement of the witness. The answer comes directly from the metrological nature \cite{Girolami2014,Adesso2014} of our non-Markovianity indicator. In optical interferometry \cite{Rafau2015}, a linear scaling of the quantum Fisher information with the mean energy per mode $\bar{n}$ denotes the shot noise, or standard quantum limit: separable probes can never surpass this limit. Conversely, entangled probes can allow for an enhanced estimation, namely a scaling of the quantum Fisher information proportional to $\bar{n}^2$, which is the so-called Heisenberg limit \cite{escher2011,Rafau2015}. Given that our witness $\mathcal{N}_Q^{\boldsymbol\sigma}$ is constructed directly from the quantum Fisher information, we may adopt the same metrological terminology in order to characterize the efficiency of our non-Markovianity witness and measure. We may then observe that the entangled STS allow for non-Markovianity detection at the Heisenberg limit, leading to a non-Markovianity measure scaling quadratically with the mean energy of the probes, while the separable MTS (and any other separable state) can still detect non-Markovianity, but providing a witness scaling at most linearly with the mean energy. We have compared the witnesses derived from optimized STS and MTS in the damping master equation in Fig.~\ref{damping3} as a function of both $\alpha$ and the mean energy per mode $\bar{n}$, generalizing the curves reported in Fig.~\ref{damping2}. The quadratic versus linear scaling with $\bar{n}$ is evident in the comparison, although both quantities are monotonically increasing functions of $\alpha$ and are nonzero as soon as $\alpha>0$.

\begin{figure*}[t]
\centering
{\includegraphics[width=7cm]{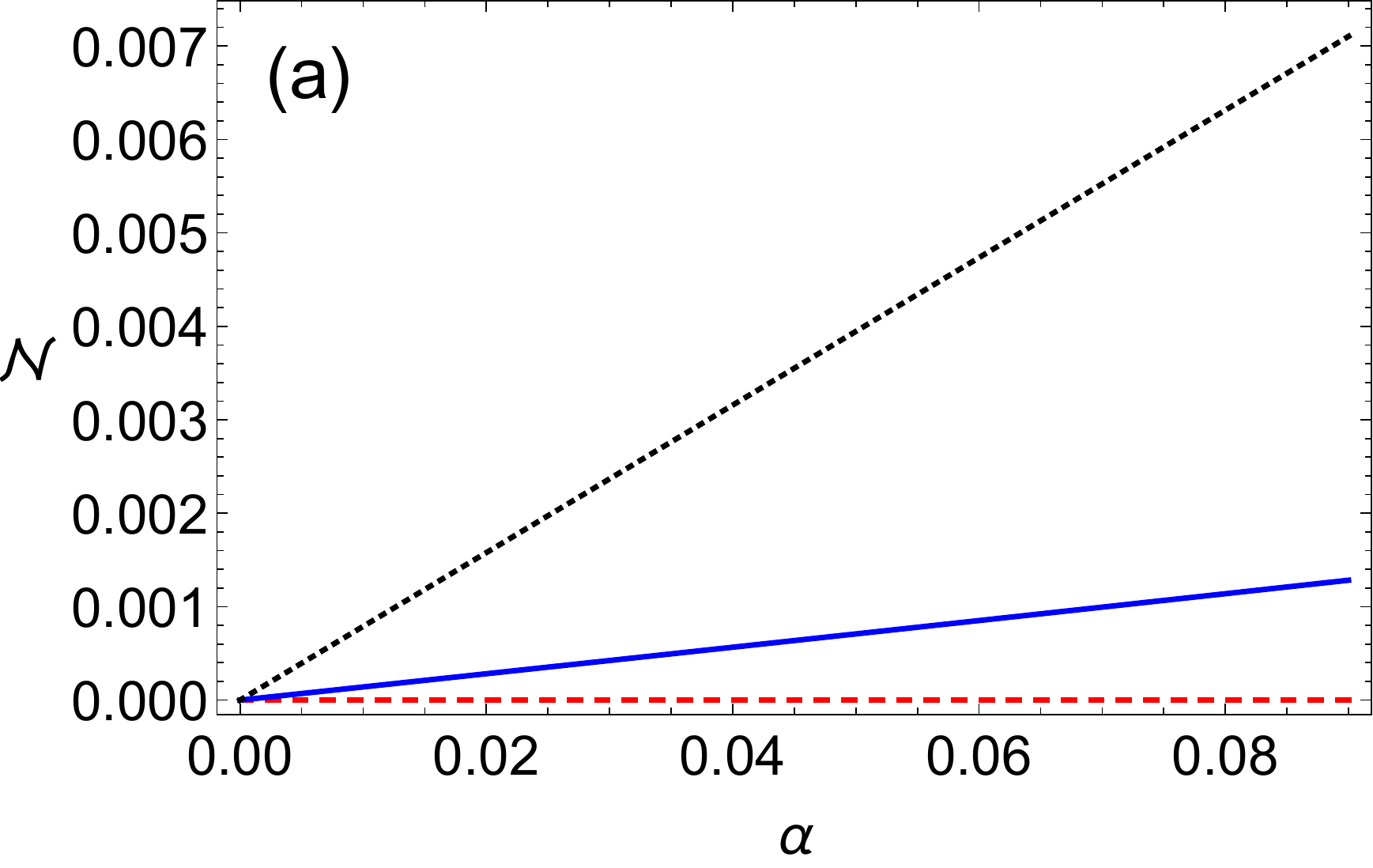}}\hspace{1cm}
{\includegraphics[width=7cm]{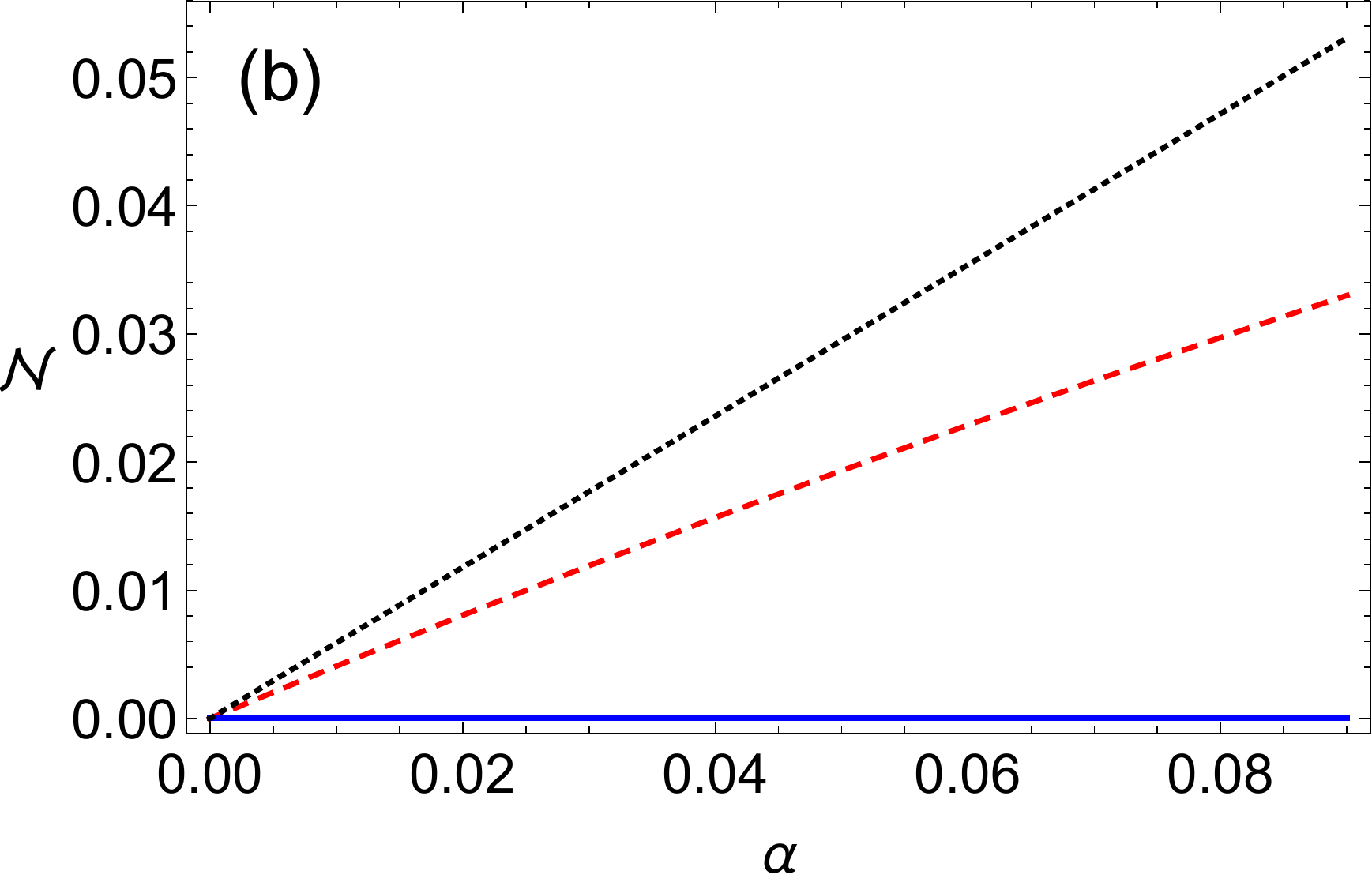}}\\[0.5cm]
{\includegraphics[width=7cm]{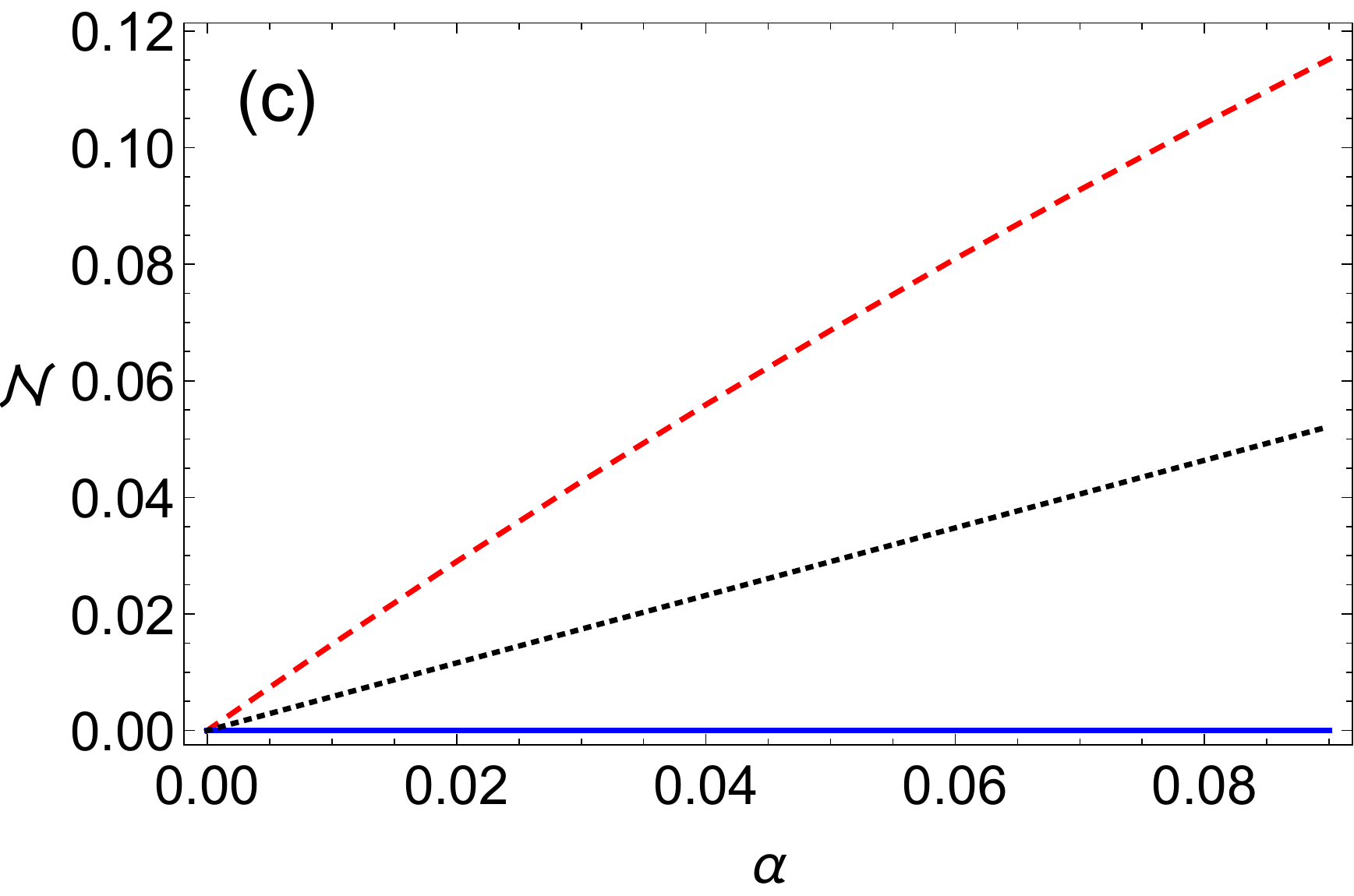}}\hspace{1cm}
{\includegraphics[width=7cm]{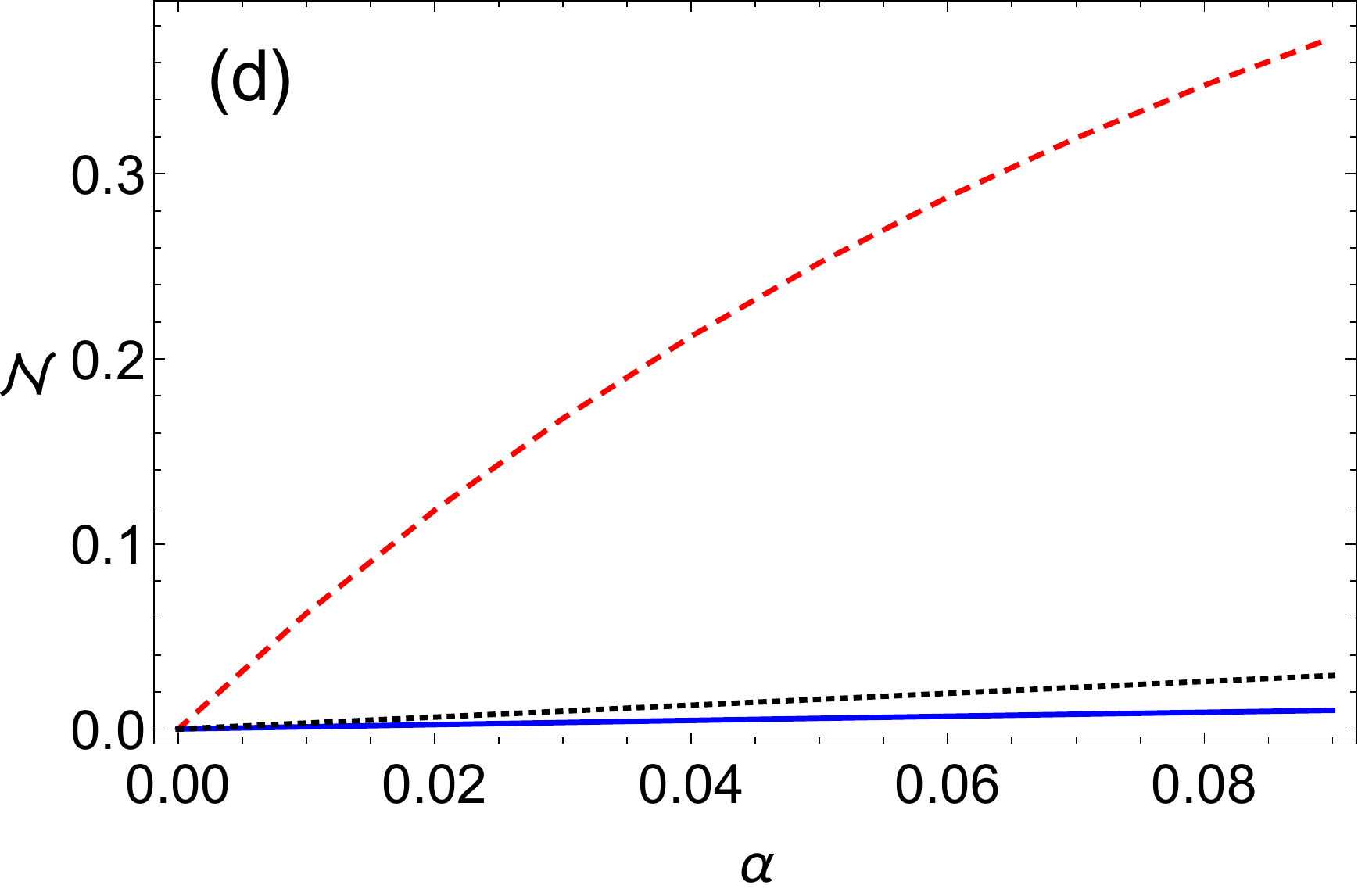}}
\caption{(Color online) Non-Markovianity indicators in the quantum Brownian motion, plotted as a function of $\alpha$, for different (scaled) reservoir temperatures: (a) $T = 0$, (b) $T = 0.5$, (c) $T = 1$, and (d) $T = 4$. The non-Markovianity witness $\mathcal{N}_Q^{\boldsymbol\sigma}$ is plotted for the squeezed thermal state with $k_2=1$ (red online dashed lines) and the mixed thermal state with $k_1=1$ (blue online solid lines). All the states have fixed average number of excitations, $\bar{n} = 2.5$. The Ohmic cut-off frequency and the characteristic frequency have been set at $\omega_c=1$ and $\omega_0=4$, respectively. The measure based on divisibility criteria, $\mathcal{N}_D$, is plotted as well for comparison (black online dotted lines). All the quantities plotted are dimensionless.}
\label{QBM4}
\end{figure*}

\subsection{Non-Markovianity in the quantum Brownian motion}

In this section we consider the non-Markovian dynamics of Gaussian states under the master equation for quantum Brownian motion (QBM). Such a dynamics has been recently observed experimentally in an opto-mechanical setup \cite{eisert2015}. Using the interaction picture, with a secular and weak coupling approximation, the master equation for QBM acting on the system mode $A$ is given by \cite{maniscalo2004},
\bq
\frac{d \rho}{dt} &=& \alpha \frac{\Delta(t) + \gamma(t)}{2} (2 \hat{a} \rho \hat{a}^\dagger - \{ \hat{a}^\dagger \hat{a}, \rho \}_+) \nonumber \\ && + \alpha \frac{\Delta(t) - \gamma(t)}{2} (2 \hat{a}^\dagger \rho \hat{a} - \{ \hat{a} \hat{a}^\dagger, \rho \}_+), \label{QBM}\eq
where $\alpha$ is the coupling constant, and the coefficients $\Delta(t)$ and $\gamma(t)$ are the diffusion and damping term, respectively. 
The master equation for QBM is an approximation of the exact master equation \cite{vasile2011, Hu1992},
\bq \frac{d \rho}{dt} &=& - \Delta(t) [\hat{q}_A,[\hat{q}_A, \rho]] + \Pi(t) [\hat{q}_A,[\hat{p}_A,\rho]] + \nonumber \\ && +\frac{i}{2} R(t) [\hat{q}_A^2,\rho] - i \gamma(t) [\hat{p}_A,\{\hat{p}_A,\rho\}_+],
\eq
where the time-dependent coefficients appearing in the equation depend on the the reservoir spectral density, and are interpreted as follows: $R(t) \rightarrow$ phase shift; $\gamma(t) \rightarrow$ damping; $\Delta(t)\ (\Pi(t)) \rightarrow$ normal (anomalous) diffusion.
The diffusion and damping coefficients in the  master equation given by Eq.~\eqref{QBM} can be written as
\be \Delta(t) = \int_0^t ds \int_0^\infty d \omega J(\omega) \left( N(\omega) + \frac{1}{2} \right) \cos (\omega_0 s) \cos(\omega s); \label{deltat} \ee
\be \gamma(t) = \int_0^t ds \int_0^\infty d \omega J(\omega) \sin(\omega_0 s) \sin(\omega s), \label{gammat}\ee
where $N(\omega) = (\exp[\hbar \omega/k_B \mathcal{T}] -1)^{-1}$ is the mean number of thermal photons with frequency $\omega$, and $\omega_0$ is the characteristic frequency of the system. We note that the coefficients $\Delta(t)$ and $\gamma(t)$ can be derived once the spectral density $J(\omega)$ and temperature $\mathcal{T}$ of the reservoir are known. In our investigation, we consider the specific case of an Ohmic spectral density of the reservoir with an exponential cut-off frequency $\omega_c$, such that $J(\omega) = \omega \ei^{- \omega/\omega_c}$.

\begin{figure*}[t]
\centering
{\includegraphics[width=7cm]{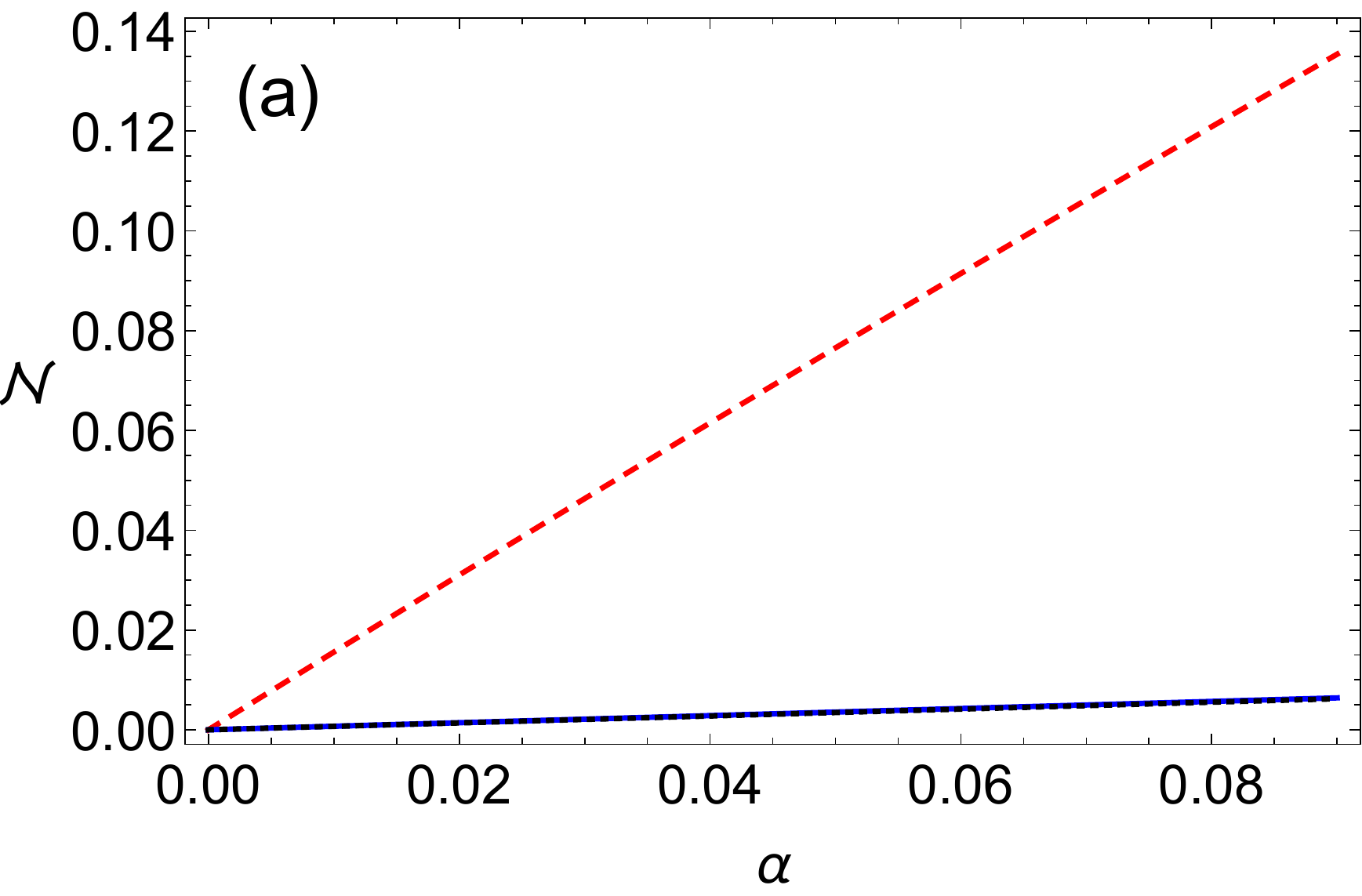}}\hspace{1cm}
{\includegraphics[width=7cm]{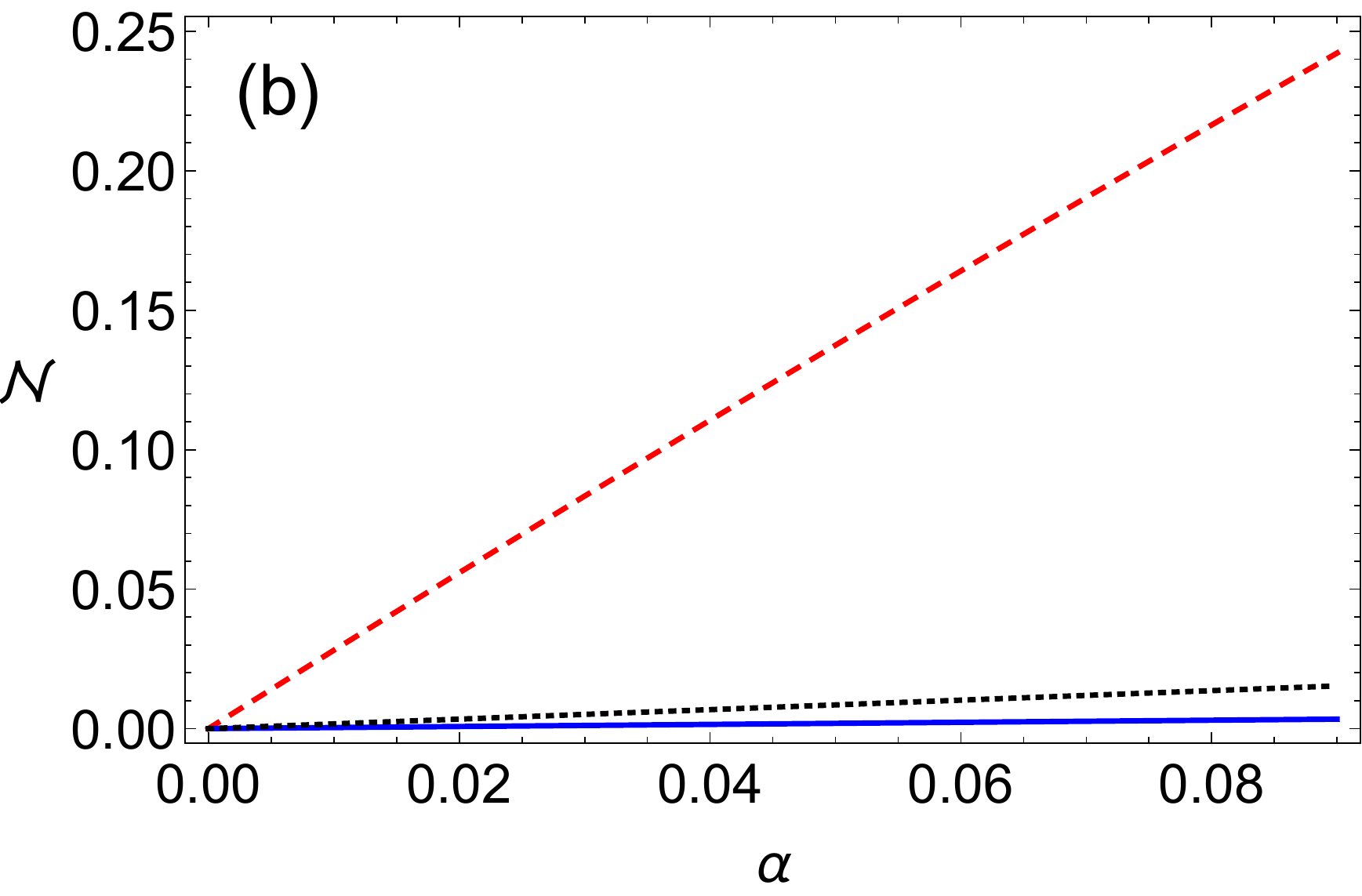}}\\[0.5cm]
{\includegraphics[width=7cm]{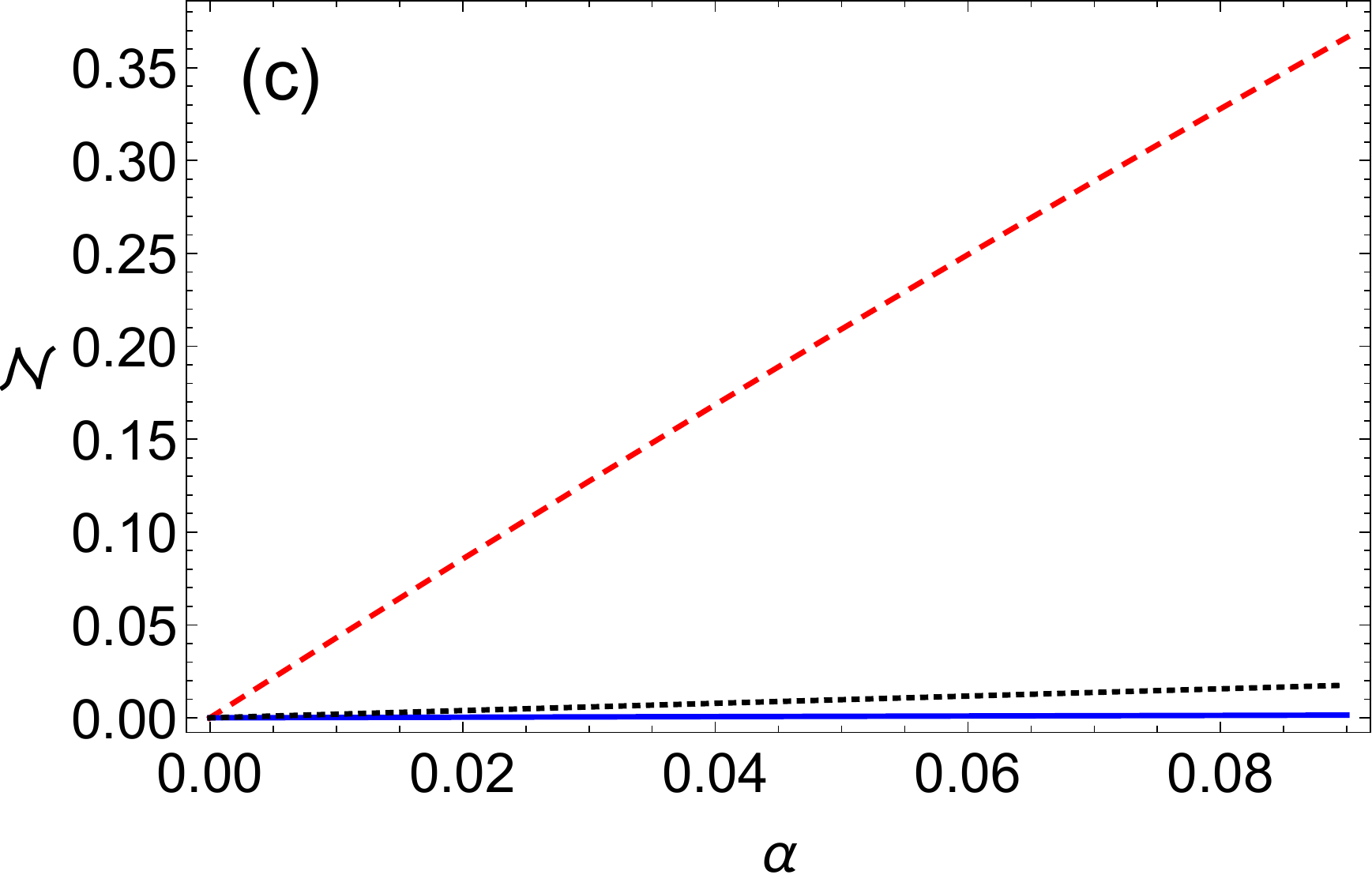}}\hspace{1cm}
{\includegraphics[width=7cm]{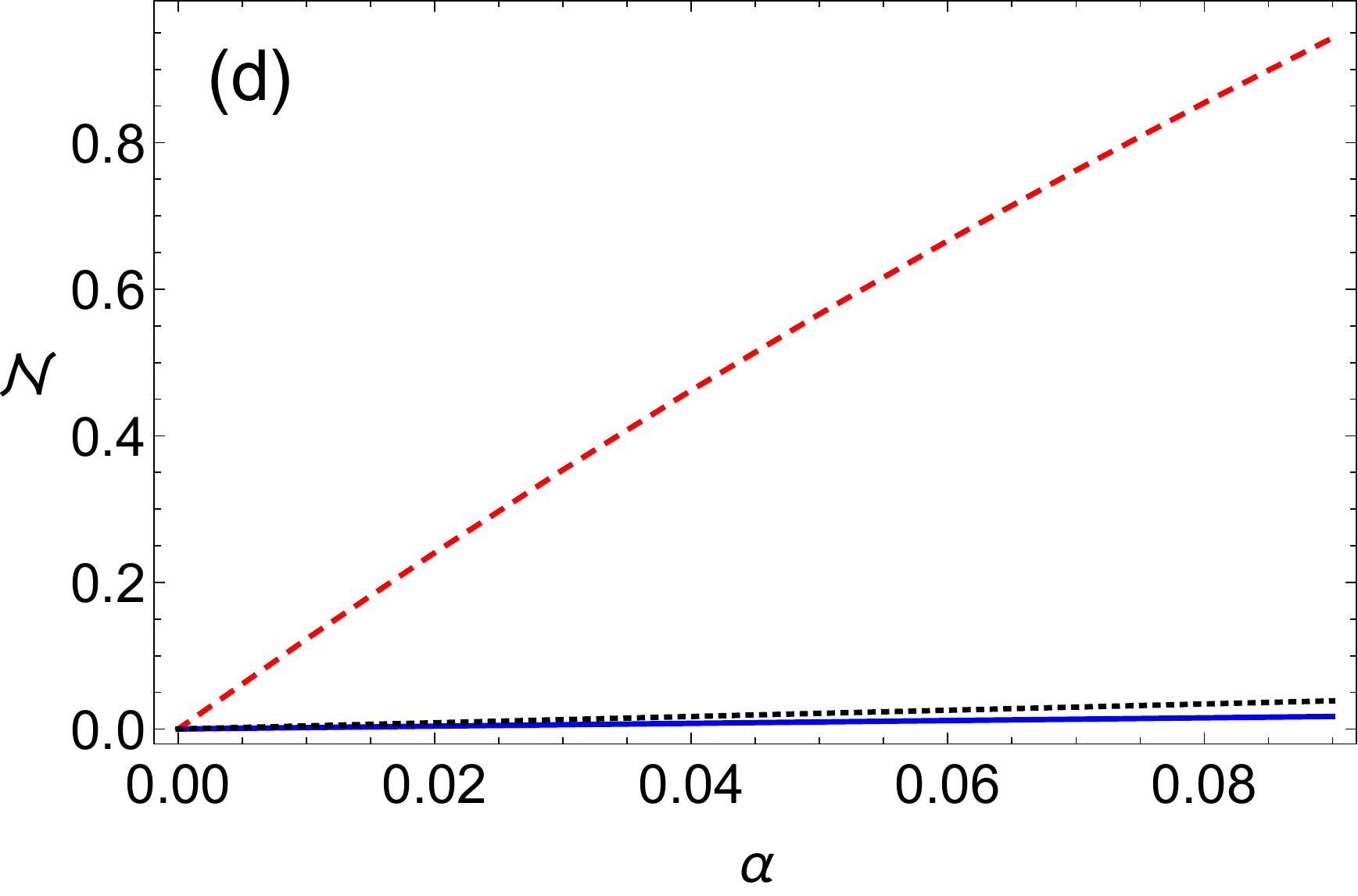}}
\caption{(Color online) Non-Markovianity indicators in the quantum Brownian motion, plotted as a function of $\alpha$, for different (scaled) reservoir temperatures: (a) $T = 0$, (b) $T = 0.5$, (c) $T = 1$, and (d) $T = 4$. The non-Markovianity witness $\mathcal{N}_Q^{\boldsymbol\sigma}$ is plotted for the squeezed thermal state with $k_2=1$ (red online dashed lines) and the mixed thermal state with $k_1=1$ (blue online solid lines). All the states have fixed average number of excitations, $\bar{n} = 2.5$. The Ohmic cut-off frequency and the characteristic frequency have been set at $\omega_c=1$ and $\omega_0=6$, respectively. The measure based on divisibility criteria, $\mathcal{N}_D$, is plotted as well for comparison (black online dotted lines). All the quantities plotted are dimensionless.}\label{QBM5}
\end{figure*}

Considering as usual an initial two-mode covariance matrix $\boldsymbol{\sigma}_{AB}(0)$, and letting mode $A$ undergo a QBM evolution under the master equation given by Eq.~\eqref{QBM}, the evolved covariance matrix is obtained as follows:
\bq \boldsymbol\sigma_{AB}(t) &=& [(\ei^{-x(t)/2} \mathbb{I}_A) \oplus \mathbb{I}_B]^T \boldsymbol\sigma_{AB}(0) [(\ei^{-x(t)/2} \mathbb{I}_A) \oplus \mathbb{I}_B] \nonumber \\ && + \alpha \ei^{-x(t)} \int_0^t \ei^{x(s)} \Delta(s) ds \left( \mathbb{I}_A \oplus \mathbb{O}_B \right),
\label{eqqbm}
\eq
where $x(t) = \alpha \int_0^t 2 \gamma(s) ds$, and the expressions of $\Delta(t)$ and $\gamma(t)$ are given by Eqs.~\eqref{deltat} and \eqref{gammat}, respectively.
The dynamics of the Gaussian state under local QBM, given via Eq.~\eqref{eqqbm}, is  mathematically more intricate, as compared to the case of Eq.~\eqref{mapsigma1} for the local damping channel, due to the presence of two decay components, $\Delta(t)$ and $\gamma(t)$. The non-Markovian dynamics of the Gaussian system is determined by the interplay between these two terms. It is known that, for the QBM master equation, the non-divisibility condition given by Eq.~\eqref{c-div} is satisfied for $\Delta(t)<|\gamma(t)|$, or equivalently for negative values of $\frac{\Delta(t)+\gamma(t)}{2}$ and $\frac{\Delta(t)-\gamma(t)}{2}$ \cite{vasile2011,torre2015}.

For the specific Ohmic spectral density under consideration, the non-Markovianity in the QBM depends on the ratio $\omega_0/\omega_c$ between the characteristic frequency of the system, $\omega_0$, and the cut-off frequency, $\omega_c$. In particular, one can expect that in the regime ${\omega_0}/{\omega_c} \ll 1$, for any temperature $\mathcal{T}$, the dynamics of the system is essentially Markovian \cite{torre2015, maniscalo2004}, as the characteristic time of the system is much larger than the corresponding relaxation time of the bath. In contrast, the regime ${\omega_0}/{\omega_c} > 1$, where the characteristic time of the system is smaller or comparable to the bath relaxation time, is more akin to physically produce non-Markovian behavior. Moreover, we note that, as the evolution time tends to infinity, all non-Markovian signatures of the Gaussian channel are lost \cite{torre2015}, irrespective of the system and bath parameters. For finite evolution times, the interplay between the diffusion and damping coefficients in determining the non-Markovian dynamics, in the regime ${\omega_0}/{\omega_c} \gtrsim 1$, is dependent on the temperature $\mathcal{T}$ of the reservoir. In the high-$\mathcal{T}$ limit, ${\Delta(t)} \gg {\gamma(t)}$, and the non-Markovianity of the evolution is solely determined by the diffusion coefficient; specifically by the condition ${\Delta(t)} < 0$ \cite{torre2015,vasile2011,maniscalo2004}. However, in the low-$\mathcal{T}$ regime, both ${\Delta(t)}$ and ${\gamma(t)}$ are comparable and the dynamics is intrinsically richer. In our investigation of non-Markovianity in QBM, we analyze both the low and high temperature regime, with the ratio ${\omega_0}/{\omega_c}$ greater than unity. Furthermore, motivated by the analysis in the previous section, we only consider the non-Markovian witness $\mathcal{N}_Q^{\boldsymbol\sigma}$ evaluated for the two classes of initial Gaussian states represented by Eqs.~\eqref{sigmamixed} and \eqref{sigmasqueezed}, with $k_1=k_2=1$.

In Figs.~\ref{QBM4} and \ref{QBM5}, we have plotted $\mathcal{N}_Q^{\boldsymbol\sigma}$
for values of the scaled temperature $T$ (= $\frac{k_B \mathcal{T}}{\hbar \omega_c}$) ranging from 0 (vanishing) to 4 (high), with the ratio ${\omega_0}/{\omega_c}$ equal to 4 and 6, respectively. We numerically calculated the non-Markovian witness for the two classes of initial Gaussian states, namely the entangled squeezed thermal state and the separable mixed thermal state. For comparison, we also plot  in Figs.~\ref{QBM4} and \ref{QBM5} the non-Markovian measure constructed from the violation of divisibility, $\mathcal{N}_D$, given by Eq.~\eqref{divisibility} \cite{torre2015}.
The non-Markovian behavior of the channel for QBM is evident in these figures, for both indicators $\mathcal{N}_D$ and $\mathcal{N}_Q^{\boldsymbol\sigma}$, obtained via non-divisibility and using GIP, respectively.
We note that, apart from the case of small ratio ${\omega_0}/{\omega_c}$  at $T = 0$ case, the entangled two-mode STS with $k_2 = 1$ appears to give the maximum value of the non-Markovian witness among the considered initial probes. However, the separable two-mode MTS  with $k_1 = 1$ consistently provides a finite value of  $\mathcal{N}_Q^{\boldsymbol\sigma}$ in a broad parameter range, thus demonstrating once more that entanglement between the two modes in the input probe states is not necessary in principle to detect non-Markovianity of Gaussian channels, specifically in the case of the QBM. Recall that this was explicitly shown for the damping master equation in Sec.~\ref{beyent} as well.

Quite surprisingly, we observe that at $T = 0$, the witness $\mathcal{N}_Q^{\boldsymbol\sigma}$ constructed from the entangled pure STS, in the regime ${\omega_0}/{\omega_c} = 4$, does not show any perceptible value as compared to the one constructed from the separable MTS, or to the measure $\mathcal{N}_D$. Therefore, in this case, mixed separable states are necessary in some cases to detect non-Markovianity of zero-temperature QBM whereas entangled probes would fail to detect it, within our approach. A possible explanation for such a phenomenon may be gained by reaching a clearer understanding of the interplay between the diffusion and damping coefficients in this regime. At very low $T$, up to first order corrections, the diffusion coefficient $\Delta(t)$ is independent of $T$ and does not uniquely dominate non-Markovian characteristics in regimes close to resonance, ${\omega_0}\approx{\omega_c}$ \cite{vasile2011}. Hence, at $T$ = 0, the non-Markovian behavior is primarily dependent on the value of $\gamma(t)$ with respect to  $\Delta(t)$. This regime of imperceptible $\Delta(t)$ dependence, is not captured by the pure STS  ($k_2 = 1$) in Fig.~\ref{QBM4}.
However, as the ratio ${\omega_0}/{\omega_c}$ becomes higher and gets away from resonance, $\Delta(t)$ becomes negative and dominates the non-Markovian behavior, which is now duly captured by the pure STS in Fig.~\ref{QBM5}.
%

\begin{figure}[h]
\begin{flushleft}  
{\includegraphics[height=7cm]{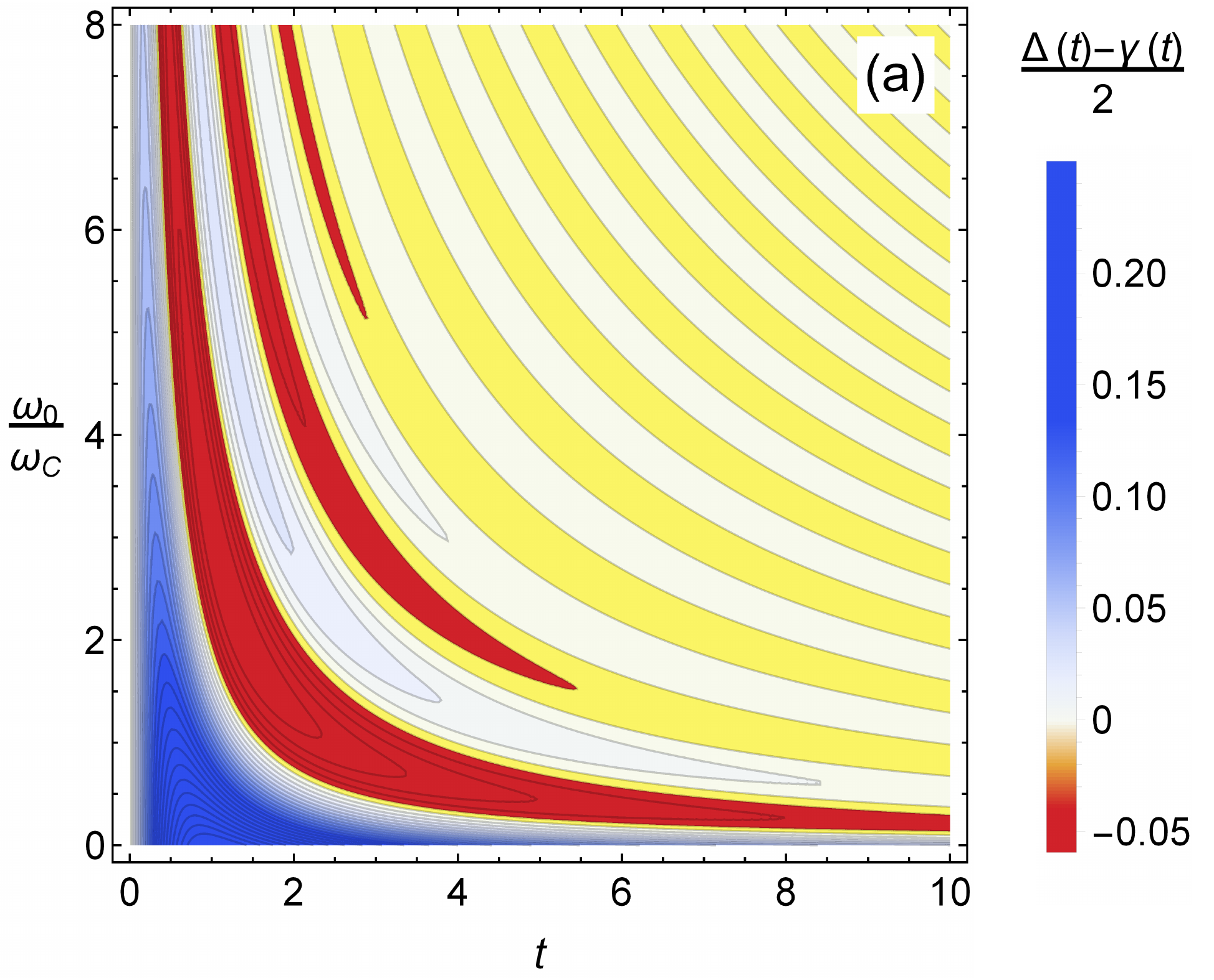}}\\[0.3cm]
{\includegraphics[height=7cm]{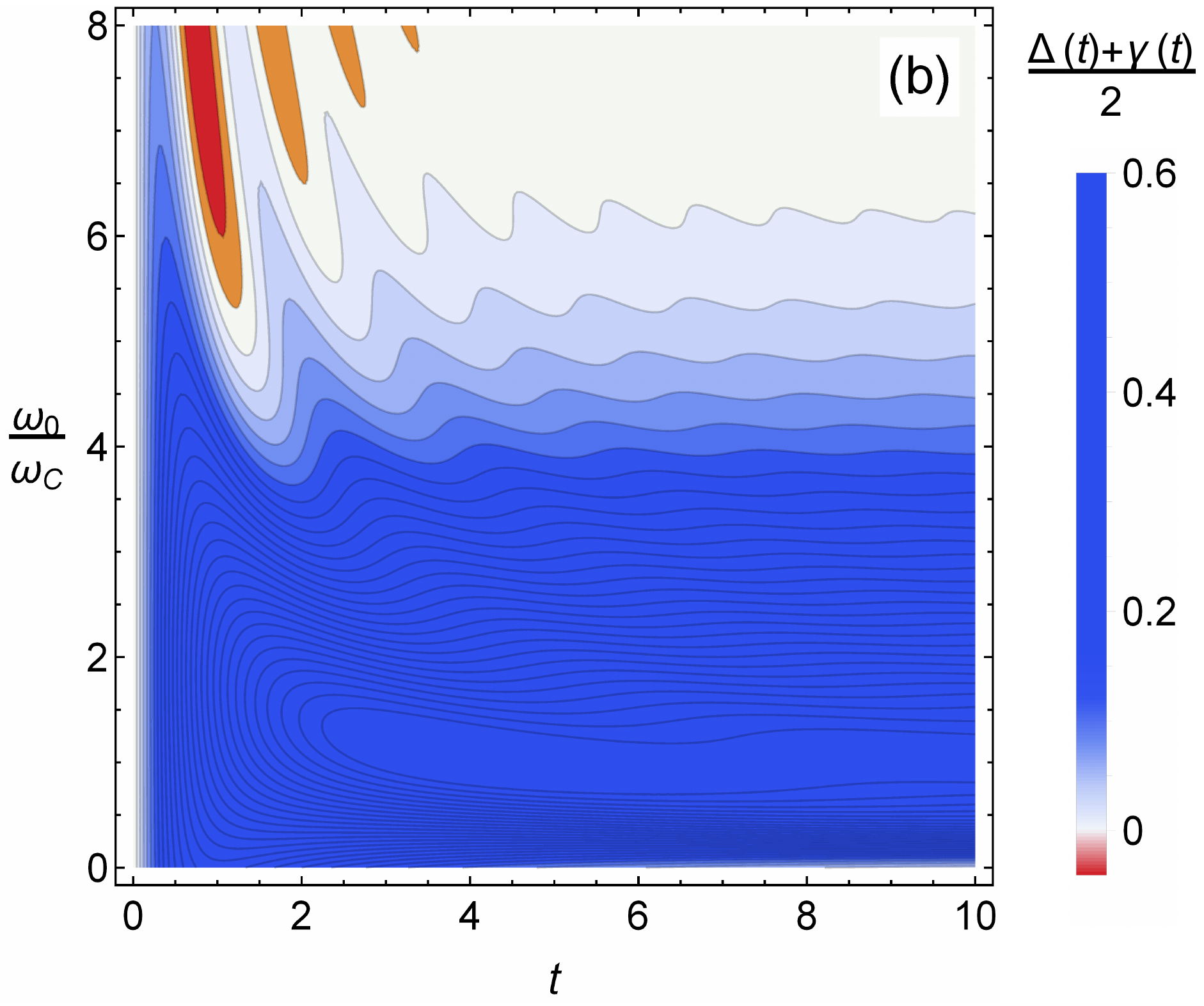}}
\end{flushleft}
  \caption{(Color online) We plot the temporal evolution of the quantities (a) $(\Delta(t) - \gamma(t))/{2}$ and (b) $(\Delta(t) + \gamma(t))/{2}$, under the master equation for the quantum Brownian motion, in the $T = 0$ regime, as a function of time $t$ and of the ratio $\omega_0/\omega_c$. Notice how the quantity in (a) has several negative oscillations, while the quantity in (b) is only negative in a small region at short time and high frequency ratio. All the quantities plotted are dimensionless.}\label{QBM6}
\end{figure}

%
In Fig.~\ref{QBM6} we depict the interplay between damping and diffusion at $T=0$, giving a pictorial exposition of the arguments presented above. We  plot specifically  the quantities $(\Delta(t) + \gamma(t))/{2}$ and $(\Delta(t) - \gamma(t))/{2}$, for $0 < {\omega_0}/{\omega_c}<8$. The figures show that the quantity $(\Delta(t) + \gamma(t))/{2}$ is always positive at the regime closer to resonance, as compared to $(\Delta(t) - \gamma(t))/{2}$ which features negative peaks. In this regime, the non-Markovianity is not dependent on ${\Delta(t)}$ \cite{vasile2011}. However, in the regime ${\omega_0}/{\omega_c} \gg 0$, both quantities $(\Delta(t) + \gamma(t))/{2}$ and $(\Delta(t) - \gamma(t))/{2}$ can be negative, with $\Delta(t)$ having a stronger impact on the non-Markovianity.

%
%

\section{Conclusion}
\label{concl}

In this work we have presented a novel approach to witness---via $\mathcal{N}_Q^{\boldsymbol\sigma}(\Lambda)$, Eq.~\eqref{witness}---and measure---via $\mathcal{N}_Q(\Lambda)$, Eq.~\eqref{nonmarkov}---the non-Markovianity of a continuous variable Gaussian channel $\Lambda$, using bipartite Gaussian states as probes. Our framework is based on the breakdown of monotonicity of an operational figure of merit defined in the context of quantum metrology,  the so-called Gaussian interferometric power \cite{Adesso2014,Bera2014}. This work and Ref.~\cite{dhar2015}, taken together, demonstrate that a general framework based on the interferometric power can be very fruitful to achieve an experimentally feasible characterization and, in some relevant cases, an analytical quantification of non-Markovianity in dynamical maps spanning from qubits to continuous variable systems.

In the present paper we have applied our methods to study in detail two examples of open Gaussian dynamics, the damping master equation and the quantum Brownian motion.

For the damping master equation, we have calculated the measure $\mathcal{N}_Q$ exactly, showing that pure entangled two-mode squeezed states are the optimal probes for the detection of non-Markovianity. Since our indicator is based on the quantum Fisher information, we may borrow a metrological terminology and argue that non-Markovianity can be detected at the Heisenberg limit, as reflected by the fact that $\mathcal{N}_Q \equiv \mathcal{N}_Q^{STS}$ exhibits a quadratic scaling with the mean energy of the probes. However, even separable states such as mixed thermal states are found to be useful to detect non-Markovianity, albeit resulting in suboptimal witnesses $\mathcal{N}_Q^{MTS}$ which scale at most linearly with the mean energy of the probes.

For the quantum Brownian motion, we have analyzed different regimes in terms of the reservoir temperature and the bare frequency of the system. We have shown that our approach successfully witnesses non-Markovianity for all regimes in which this behavior is expected to manifest, and our results are consistent with a recent characterization of non-Markovianity in terms of the non-divisibility criterion for Gaussian channels \cite{torre2015}. It is worth mentioning that we found some particular regimes (at zero temperature) where non-Markovianity is better detected using initial states which are mixed and with no entanglement, as compared to pure two-mode squeezed probes.

Overall, our approach demonstrates that cheap (in terms of engineering demands) two-mode resource states can be adopted to construct robust non-Markovianity witnesses for Gaussian  channels acting on one mode of the system.

A complete and practical characterization of non-Markovianity in continuous variable open systems, specifically tailored to Gaussian states and Gaussian channels, may be of great importance for technological applications \cite{breuer2015}, especially if one is interested in using memory effects and backflow of information from the environment,  that can arise from non-Markovian evolutions, to protect and enhance quantum information tasks such as communication, cryptography and metrology. The role of non-Markovianity is also very promising in quantum thermodynamics, since non-Markovian channels can, in principle, allow work extraction accompanied by an increase of the system entropy \cite{Bylicka2015}. We believe that our approach can be further developed and applied to assess environmental enhancements in quantum technologies realized with continuous variable systems, as well as to reach a more fundamental understanding of the physical mechanisms underpinning non-Markovianity in different realizations of Gaussian quantum channels.

\begin{acknowledgments}
L.A.M.S. acknowledges financial support from Brazilian Agencies CAPES (6842/2014-03) and CNPq (470131/2013-6). H.S.D. and M.N.B. acknowledge financial supports from the Department of Atomic Energy, Government of India. H.S.D. also thanks the University of Nottingham for hospitality and support during a visit. M.N.B. also acknowledges funding from Spanish project FOQUS (FIS2013-46768), 2014 SGR 874, and the John Templeton Foundation.
P.L.-S. acknowledges financial support from the University of Nottingham via a Graduate School Travel Prize. G.A. acknowledges financial support from the Foundational Questions Institute (Physics of Information Grant No. FQXi-RFP3-1317) and the European Research Council (ERC StG GQCOP Grant No.~637352).
\end{acknowledgments}


%

\end{document}